\documentclass{elsart}
\usepackage{graphicx,natbib,appendix}
\journal{New Astronomy}
\newcommand{\beq}{\begin{equation}}
\newcommand{\eeq}{\end{equation}}
\newcommand{\beqarray}{\begin{eqnarray}}
\newcommand{\eeqarray}{\end{eqnarray}}

\def\lsim{\raise0.3ex\hbox{$\;<$\kern-0.75em\raise-1.1ex\hbox{$\sim\;$}}}
\def\gsim{\raise0.3ex\hbox{$\;>$\kern-0.75em\raise-1.1ex\hbox{$\sim\;$}}}
\def\half{\frac{1}{2}}

\def\cm{\,{\rm cm}}
\def\km{\,{\rm km}}

\def\s{\,{\rm s}}

\def\pc{\,{\rm pc}}
\def\kpc{\,{\rm kpc}}

\def\gev{\,{\rm GeV}}

\def\kmps{\km\s^{-1}}
\def\msun{\,{\rm M}_\odot}
\def\rhodm{\rho_{\rm\scriptscriptstyle DM}}
\def\rhodmsun{\rho_{{\rm\scriptscriptstyle DM},\odot}}
\def\phidm{\phi_{\rm\scriptscriptstyle DM}}
\def\vdispsq{\langle v^2 \rangle}
\def\vdisp{\langle v^2 \rangle^{1/2}}
\def\vrdispsq{\langle v_r^2 \rangle}
\def\vrdisp{\langle v_r^2 \rangle^{1/2}}
\def\vcdispsq{\langle v_c^2 \rangle}
\def\vcdisp{\langle v_c^2 \rangle^{1/2}}
\def\vtdispsq{\langle v_t^2 \rangle}

\def\vdispdm{\vdisp_{\rm\scriptscriptstyle DM}}
\def\vdispdmsun{\vdisp_{{\rm\scriptscriptstyle DM},\odot}}
\def\rmin{r_{\rm min}}
\def\rmax{r_{\rm max}}
\def\rminsq{r^2_{\rm min}}
\def\rmaxsq{r^2_{\rm max}}
\def\vr{v_r}
\def\vt{v_t}

\def\vrmax{v_{r,{\rm max}}}
\def\vtmin{v_{t,{\rm min}}}
\def\vtmax{v_{t,{\rm max}}}

\def\vrmaxsq{v^2_{r,{\rm max}}}
\def\vtminsq{v^2_{t,{\rm min}}}
\def\vtmaxsq{v^2_{t,{\rm max}}}
\begin{document}
\begin{frontmatter}
\title{Dynamics of dwarf-spheroidals and the dark matter halo of the Galaxy}
\vspace{-0.5cm}
\author[wustl,iia]{R. Cowsik},
\ead{cowsik@wuphys.wustl.edu}
\author[iia]{Charu Ratnam\thanksref{now}},
\thanks[now]{Deceased}
\author[iia,sinp]{Pijushpani Bhattacharjee\corauthref{cor}},
\corauth[cor]{Corresponding author.}
\ead{pijush.bhattacharjee@saha.ac.in}
\author[cita]{Subhabrata Majumdar\thanksref{subha_now}}
\thanks[subha_now]{Now at Tata Institute of Fundamental Research, Homi 
Bhabha Road,\\
Mumbai 400005. India}
\ead{subha@tifr.res.in}
\address[wustl]{McDonnell Center for the Space Sciences and Department of
Physics, Washington University, St.~Louis, MO 63130, USA}
\address[iia]{Indian Institute of Astrophysics, Koramangala, Bangalore
560034, India}
\address[sinp]{Theory Division, Saha Institute of Nuclear Physics, 1/AF  
Bidhan Nagar,\\
Kolkata 700064, India}
\address[cita]{CITA, University of Toronto, 60 St George St, Toronto,
Ontario, Canada}
\begin{abstract}
The dynamics of the dwarf spheroidal (dSph) galaxies in the 
gravitational field of the Galaxy is investigated with particular 
reference to their susceptibility to tidal break-up. Based on the observed 
paucity of the dSphs at small Galactocentric distances, we put forward the 
hypothesis that subsequent to the formation 
of the Milky Way and its satellites, those dSphs 
that had orbits with small perigalacticons were tidally
disrupted, leaving behind a population that now has a 
relatively larger value of its average perigalacticon to apogalacticon 
ratio and consequently a larger value of its r.~m.~s. transverse to radial 
velocities ratio compared to their values at the time of formation of the 
dSphs. We analyze the implications of this hypothesis for the phase
space distribution of the dSphs and that of the dark matter (DM)
halo of the Galaxy within the context of a self-consistent model in which
the functional form of the phase space distribution of DM particles 
follows the King model i.e. the `lowered isothermal' distribution and 
the potential of the Galaxy is determined self-consistently by
including the gravitational cross-coupling between visible matter and DM
particles. This analysis, coupled with virial arguments, 
yields an estimate of $\gsim 270 \kmps$ for the circular 
velocity of any test object at galactocentric distances of $\sim100\kpc$, 
the typical distances of the dSphs. The corresponding self-consistent 
values of the relevant DM halo model parameters, 
namely, the local (i.e., the solar neighbourhood) values of the DM 
density and velocity dispersion in the King model and its truncation 
radius, are estimated to be $\sim 
0.3\gev/\cm^3$, $>350\kmps$ and $\gsim 150\kpc$, respectively. 
Similar self-consistent studies with other possible forms of the DM 
distribution function 
will be useful in assessing the robustness of our estimates of the 
Galaxy's DM halo parameters. 
\end{abstract}
\vspace{-0.5cm} 
\end{frontmatter}
\section{Introduction
\label{sec:intro}}
A well-motivated conjecture is that electrically neutral weakly
interacting particles generated in the big-bang origin of the
Universe through their gravitation triggered the formation of
galaxies \citep{CM73,Einasto}. This process generically leads to the
formation of halos of dark matter (DM) surrounding the galaxies. The
DM halo in which the Milky Way Galaxy is embedded is the subject of
this paper. Detailed study of the DM halo surrounding the Galaxy is
particularly important because it might provide insights into the
more general problem of DM in the Universe and the role of DM in the
formation and dynamics of the diverse galactic systems.

In constructing theoretical models of the DM halo of the Galaxy one
must keep in mind that the visible matter in the form of stars and
gas in the Galaxy not only act as tracers of the galactic potential
but also contribute to it, dominating it at galactocentric distances
below a few kiloparsecs and diminishing in importance at very large
distances where DM is the main contributor. It is the interplay
between these two components that finally determines the spatial
distribution of both the visible and the dark matter components of
the Galaxy.

What is the mass of the Galaxy including its halo? How far does the
halo extend? These are some of the crucial questions that are
addressed in this paper using a novel method that uses the dynamics
of the dwarf-Spheroidal (dSph) galaxies in the gravitational
potential of the Galaxy together with a model of the phase space
structure of the DM halo of the Galaxy that incorporates the
gravitational potential of the visible matter as well as that of the DM
particles in a self-consistent manner. Early work on using the 
satellites of the Milky Way as tracers of its gravitational potential 
includes \citet{LB_etal_83}, \citet{LT87} and \citet{WE99} (hereafter 
WE99). 

One of the most effective probes of the gravitational potential of 
the DM halo of the
Galaxy is the behavior of its rotation curve, the circular rotation
speed as a function of the galactocentric distance $R$. However, direct
measurements of the Galactic rotation curve are available only up to
$R\sim 20\kpc$. On the other hand, a proper understanding of the
nature of the DM halo requires the knowledge of the behavior of the
rotation curve at large galactocentric distances. It is in this
context that the dynamics of dSphs, which lie beyond several
tens of kpc from the Galactic center, play an important role in the
study of the nature of the DM halo. 

The central idea and the main results of this paper can be
summarized as follows: The dSphs, comprising of mostly old
population stars with total visible masses in the range $\sim 10^5$
-- $10^7\msun\,$ and radii in the range 1 -- 2 kpc, populate a
region extending from $\sim 60\kpc$ to $250\kpc$ from the Galactic
center. They are relatively low density systems that are susceptible
to tidal break-up in the gravitational potential of the Galaxy. Even
with the additional mass provided by the DM in these systems they
are near the threshold of tidal break-up in the gravitational field
of the Galaxy \citep{AARONSON83,GR94,LB_etal_83,IH95,DC99}. During the 
early epochs following
the formation of the Galaxy and its satellites, a subset of the
dSphs that had close perigalactic passages must have been tidally
disrupted leaving behind a population which do not approach the
Galaxy too closely. Thus, the present day phase space distribution
of the dSphs should be such as to be depleted of orbits with large
radial velocities, $v_r$, which would have brought them closer to the 
centre of the Galaxy than a certain minimum galactocentric 
distance $\rmin$. This makes their velocity distribution skewed in favor
of transverse velocities, $v_t$. The velocity skewness or 
anisotropy parameter, $j$, can in general be defined as 
\begin{equation}
j\equiv \vdispsq/\vrdispsq\,, \label{j_def1}
\end{equation}
where $v$ is the total velocity of a satellite and $v_r$ is its radial 
component, and the angular brackets denote the 
average over a population of these satellites. This parameter can be   
theoretically calculated for dSphs as a function of the parameter $\rmin$ 
that parametrizes the spatial restrictions imposed on the allowed orbits 
of the dSphs for an assumed form of the phase space 
distribution function of the dSphs and a given potential of the Galaxy. 
A comparison of the calculated radial distribution of the dSphs with 
the observed one then allows us to determine the most likely value of the 
parameter $\rmin$ and thereby obtain an estimate of the velocity 
anisotropy parameter $j$ of the dSphs. This, coupled with their observed 
radial velocities, then gives us an estimate of $\vdispsq$ of the dSphs.  
This in turn allows us to infer, using the virial
theorem as formulated by \citet{LF81}, the circular rotation speed
$v_c$ of any tracer object in the potential of the Galaxy at
galactocentric distances spanned by the dSphs. This analysis yields
an estimate of $v_c$ of $\sim270\kmps$ at galactocentric distances
of $\gsim100\kpc$. 

Our dynamical model for the gravitational potential of the Galaxy 
including its 
DM halo in which the dSphs move is based on simple assumptions about the 
phase space structure of the DM particles, and incorporates the 
gravitational potential of the visible matter as well as that of the DM 
particles in a self-consistent manner. The phase space 
distribution function (DF) of the finite-sized DM halo model we 
assume in this paper is the truncated or ``lowered'' isothermal DF (``King 
model'') \citep[see, e.g.,][p.~232]{BT87}. This model has three 
parameters, namely, the density, $\rhodmsun$, and velocity dispersion, 
$\vdispdmsun$, of the DM particles at the solar location, and the radius, 
$r_t$, of the DM halo. We determine these model parameters by 
demanding that the model
not only yield the observed spatial distribution of the dSphs but also at 
the same time be consistent with the directly
observed rotation speeds of the Galaxy at all galactocentric 
distances. This analysis yields good fit to 
all the available observational data for values of the King model DM 
halo parameters, $\rhodmsun\sim 0.25$ -- $0.4\gev\cm^{-3}$, 
$\vdispdmsun\gsim 350\kmps$, and $r_t\gsim 150\kpc$. 

This paper is organized as follows: 
We begin, in section \ref{sec:spatial_df_sats}, by reviewing the spatial 
distribution of the satellites and pointing out several of its relevant 
features. In section \ref{sec:truncated_df_dsph} we derive the 
constraint equations pertaining to orbits which are confined between 
certain minimal and maximal galactocentric distances in the galactic 
potential and then discuss how these constraints influence the
distribution of dSphs in phase space. In section \ref{sec:virial} we 
briefly 
review the virial theorem due to \citet{LF81} which allows us to estimate 
the circular rotation speed of any tracer object at large Galactocentric 
distances if the radial velocities of an ensemble of objects at those 
distances are known. Our self-consistent model for the DM halo of the 
Galaxy is described in section \ref{sec:dmhalo_model_math}. The rotation 
curves for various different values of the relevant parameters of the 
DM halo model 
are presented in section \ref{sec:rotation_curve}, where we also discuss 
how the measured rotation speeds allow us to determine one of the 
parameters of the model, namely, the DM density in the solar 
neighborhood. We then describe in section 
\ref{sec:constraints_from_dsphs} how the dynamics of the dSphs can be used 
to constrain the parameters of the DM halo model by comparing the 
theoretically 
calculated radial distribution of the dSphs with their observed radial 
distribution. The estimates of the velocity anisotropy parameter $j$ 
and the resulting lower limits on the average circular rotation speeds at 
large galactocentric distances spanned by the dSphs are obtained in  
section \ref{sec:j_estimation}. A summary of the main results of the paper 
is presented in section \ref{sec:summary}. Three appendices provide 
additional material that reinforces the underlying physical basis of the 
results obtained in the main body of the paper.   

\section {Spatial distribution of the satellites of the Galaxy 
\label{sec:spatial_df_sats}}
A careful compilation of 27 satellites of the Galaxy has been made by 
WE99, and our analysis in this paper is based on the astronomical data 
available on this sample of 27 objects~\footnote{Subsequent to completion 
of our analysis and prior to the publication of the paper, some new 
satellites of the Milky Way have been discovered (see, e.g., 
\citet{newsats}). It will be interesting to repeat the analysis presented 
here when data on a more complete set of satellites become 
available.}. 

We define the radial number distribution of the satellites through 
the relation 
\begin{equation}
N(r)dr=4\pi r^2 n(r)dr\,,
\label{N_r_dr_def} 
\end{equation}
where $n(r)$ is the spatial number density distribution. Observationally,  
$N(r)$ just represents the number of satellites in unit interval in $r$ at 
a radial distance $r$ from the Galactic centre irrespective of their 
angular coordinates. This one dimensional distribution is less subject to 
statistical uncertainties than higher dimensional distributions such as 
the phase space distribution. 

\begin{figure}
\rotatebox{270}{\includegraphics[width=4in]{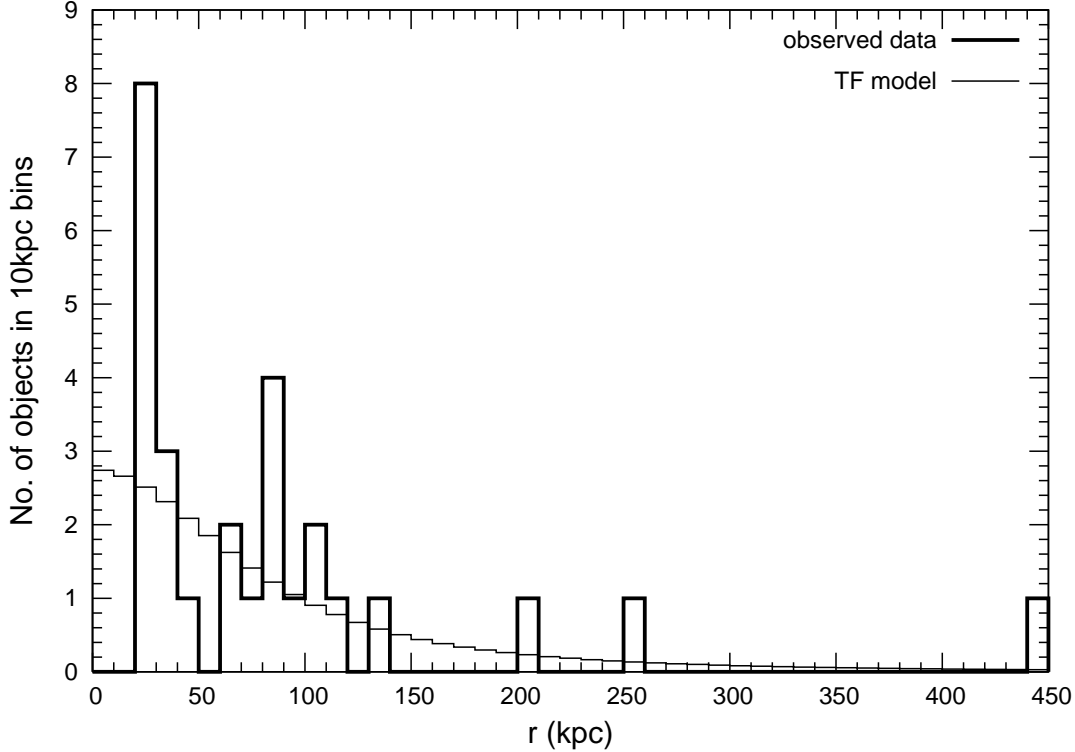}} 
\caption{Distribution of the number of satellites as a function of the 
distance from the Galactic centre. Notice the two distinct peaks 
separated by a dip at $\sim 50\kpc$. The prediction of the TF model of 
\citet{WE99} is also shown. The dwarf spheroidals all lie in the second 
peak and beyond.} 
\label{fig:hist_all_TF}
\end{figure}

We display in Fig.~\ref{fig:hist_all_TF} the observed distance 
distribution of the satellites. Even a visual inspection of this 
distribution is quite revealing: There is a sharp 
peak comprising of 12 satellites at around $30\kpc$. Then there is a 
group of 15 satellites distributed as a broad peak at around 
$90\kpc$, with a long tail extending beyond 200 kpc. On the same Figure we 
also display the best fitting ``shadow tracer'' TF model of WE99 and note 
that the sharp first peak, the dip at around 50 kpc, and the second 
peak at around 90 kpc are not well represented by the model. In fact it 
appears that no model that predicts a smooth monotonic decrease in the 
number distribution will fit the observations, and one needs to consider 
that there are two distinct populations corresponding to each of the peaks 
and separated by a low density region centered around 50 kpc. 

How can one maintain two such spatially separated populations over the age 
of the Galaxy $\sim$ 10 billion years? Unless there are dynamical 
constraints, the features such as two separate peaks and a depleted region 
in between will get smoothed out over a few orbital periods, that is, well 
within a billion years. An important clue as to the possible dynamical 
reasons for imposing these features on the distance distribution of the 
satellites stems from the fact that all the dSph galaxies lie in the 
second more distant region and none nearby. As noted in the Introduction, 
the dSphs are highly susceptible to tidal break up in the gravitational 
field of the Galaxy. The key idea is that at an early epoch when the 
Galaxy and its satellites were being formed, perhaps even prior to the 
complete condensation of the satellites into their present configurations, 
a dynamically chosen subset of the satellites were tidally disrupted by 
the centrally condensed galaxy. A fraction of the satellites might have 
been tidally stable, as would be the case for globular clusters, for 
example, which might have settled down into the distribution 
peaked at around 30 kpc that we see today.   

As for the objects populating the second peak 
around 90 kpc, it is likely that these represent a subgroup of objects,  
which even though were weekly bound at the time of their formation, 
survived tidal disruption as their orbits did not ever bring them close to 
the central galaxy. The other subgroup with more radial trajectories, on 
the other hand, were 
tidally disrupted, as their trajectories brought them much closer to the 
galaxy where the tidal fields are much stronger. Thus the tidal effects on 
this second subgroup caused  
a reduction in their densities at short galactocentric distances, with the 
concomitant effect on their phase space distribution whereby asymmetries 
were introduced into their velocity distribution. These considerations 
are relevant to any model for the formation of the satellites that 
dynamically 
removes progenitors which approached the central galaxy too closely in the 
past. We may look for this effect in the data available on the proper 
motions of six satellites given in Table 3 of WE99, three of which are 
dSphs. From these data (see also \citet{Piatek_etal_2005_2002}), it is 
straightforward to estimate the 
asymmetry parameter $j$ defined in equation (\ref{j_def1}), which yields 
$j=1+(\vtdispsq/\vrdispsq)\approx 4.7$ for all six objects and $j\approx 
4.4$ for the three dSphs, as compared to the value $j_{\rm iso}=3$ 
for an isotropic distribution of velocities.
Note, however, that because the data sample is small and uncertainties in 
the proper motion measurements sizeable the above estimate of $j$ is to 
be treated with caution, even though the data do seem to indicate the 
possible existence of a velocity anisotropy favoring 
transverse velocities 
in the distribution of the dSphs. As shown later in this paper, a more 
robust estimate of the asymmetry parameter $j$ can be obtained from the 
analysis of the number distribution $N(r)$ of the satellites without any 
reference to the data on the radial velocities nor on the proper motions 
of the satellites.   

Our $j$ parameter is related to the parameter $\beta$ defined in WE99 
through the relation 
\begin{equation}
\beta\equiv 1-\frac{\langle v_\theta^2\rangle}{\vrdispsq}
=1-\frac{\vtdispsq}{2\vrdispsq}=\frac{1}{2}(3-j)\,,
\label{j-beta-relation}
\end{equation}
which gives, for the same data set as mentioned above, $\beta\approx 
-0.85$ for all six objects and $\beta\approx -0.7$ for the three dSphs. 
Note the negative values of $\beta$ signifying possible existence of a 
velocity anisotropy in favor of transverse velocities over radial 
velocities.  

In the next section we begin by deriving the kinematic constraints on the 
orbits of dSphs and go on to show how the two observational features, 
namely, their depletion at short galactocentric distances and the velocity 
anisotropy get related to each other.    
\section {Truncated phase space of the dwarf spheroidals
\label{sec:truncated_df_dsph}}
Let us consider the motion of dSphs in the potential of the Galaxy
with its DM halo. Consider then an initial phase space distribution
function (DF) $F$ for the dSphs, from which all representative
points corresponding to orbits which do not satisfy the constraint
\begin{equation}
\rmin\leq r < \rmax 
\label{r_constraint}
\end{equation}
have been removed. Here $r$ is the radial coordinate, and $\rmin$ and
$\rmax$ are the perigalacticon and the apogalacticon of the orbits of the 
population of dSphs. For such a truncated DF we wish to evaluate the 
skewness
parameter $j$ defined in equation (\ref{j_def1}). 
\begin{figure}
\resizebox{\textwidth}{!}{\includegraphics{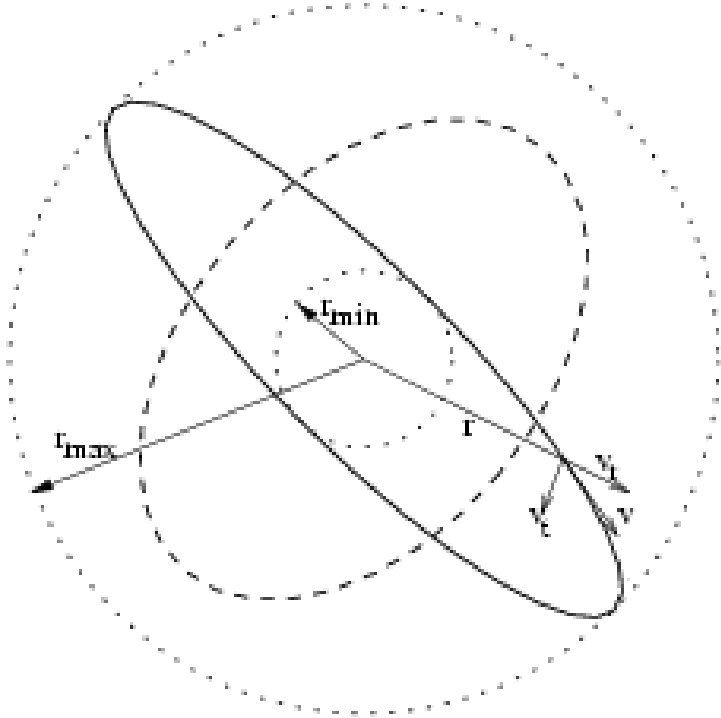}} 
\caption{The sketch shows a typical orbit (dashed curve) of 
a dwarf spheroidal galaxy around the Milky Way, and defines the various 
distances and velocities used in the calculation. It is assumed that all 
the permitted orbits of dSphs lie between $\rmin$ and $\rmax$ (see text).  
The solid curve represents the ``maximal'' orbit: This orbit has the 
maximum transverse velocity at $r_{\rm min}$ and hence permits the maximum 
radial velocity at any $r$.} 
\label{fig:dsph_orbit_fig}
\end{figure}
Now, referring to Figure~\ref{fig:dsph_orbit_fig} we can write the
equations for the conservation of energy and angular momentum of a
dSph's motion as
\begin{equation}
E = \half (v_r^2 + v_t^2) + \phi\,, \label{E_eqn}
\end{equation}
and
\begin{equation}
J = r \vt = \rmin \vt(\rmin) = \rmin\sqrt{2[E-\phi(\rmin)]}\,,
\label{J_eqn}
\end{equation}
where $E$ is the energy per unit mass, $J$ the angular momentum per
unit mass, $v_t$ the transverse velocity and $\phi$ is total
gravitational potential of the Galaxy.

Manipulating equations (\ref{E_eqn}), (\ref{J_eqn}) we can write
\begin{equation}
\vt^2(r) = \frac{\rminsq}{r^2-\rminsq} \left\{\vr^2(r) + 2[\phi(r) -
\phi(\rmin)]\right\}\,. \label{v_t_eqn}
\end{equation}
The spatial restriction imposed on the orbits by equation
(\ref{r_constraint}) implies constraints on the velocity components
$v_r$ and $v_t$, which can be deduced as follows: First, note that 
the orbit having the maximum transverse velocity at $\rmin$ will be 
the one that permits the maximum radial velocity at any $r$. Considering 
this ``maximal'' orbit,
shown schematically in Figure~\ref{fig:dsph_orbit_fig}, and using
the energy and angular momentum conservation equations given above,
it is easy to show that the maximum radial velocity, $\vrmax$, at
any $r$ is given by
\begin{equation}
\vrmaxsq(r)=2\left(\frac{\rmaxsq}{\rmaxsq-\rminsq}\right)
\left(\frac{r^2-\rminsq}{r^2}\right)\left[\phi(\rmax)-\phi(\rmin)\right]
-2\left[\phi(r)-\phi(\rmin)\right]\,.
\label{vrmaxsq_eq}
\end{equation}
The maximum $\vt$ at any $r\, $, $\vtmax\left(r,\vr(r)\right)$,
using conservation of energy, is then given by
\begin{equation}
\vtmaxsq\left(r,\vr(r)\right)=\frac{\rmaxsq}{\rmaxsq-r^2}
\left\{2\left[\phi(\rmax)-\phi(r)\right] - \vr^2(r)\right\}\,.
\label{vtmaxsq_eq}
\end{equation}
It can be shown by a Taylor expansion around $r=\rmax$, 
that this equation (\ref{vtmaxsq_eq}) correctly gives
$\vtmaxsq(r\to\rmax)=r\frac{\partial\phi}{\partial
r}|_{\rmax}=v_c^2(\rmax)$, where $v_c$ is the circular velocity. 

The minimum $\vt$ at $r\,$, $\vtmin\left(r,\vr(r)\right)$, has to be
such that the radial velocity at $\rmin$ for this orbit,
$\vr\left(\rmin\,, \vr(r)\right)\,$, vanishes, and the kinetic
energy at $\rmin$ is completely due to transverse motion. This gives
\begin{equation}
\vtminsq\left(r,\vr(r)\right)=\frac{\rminsq}{r^2-\rminsq}\left\{\vr^2(r)
+ 2\left[\phi(r)-\phi(\rmin)\right]\right\}\,. \label{vtminsq_eq}
\end{equation}
Notice, as before, $\vtminsq(r\to\rmin)=v_c^2(\rmin)$.

\begin{figure}
\rotatebox{270}{\includegraphics[width=4in]{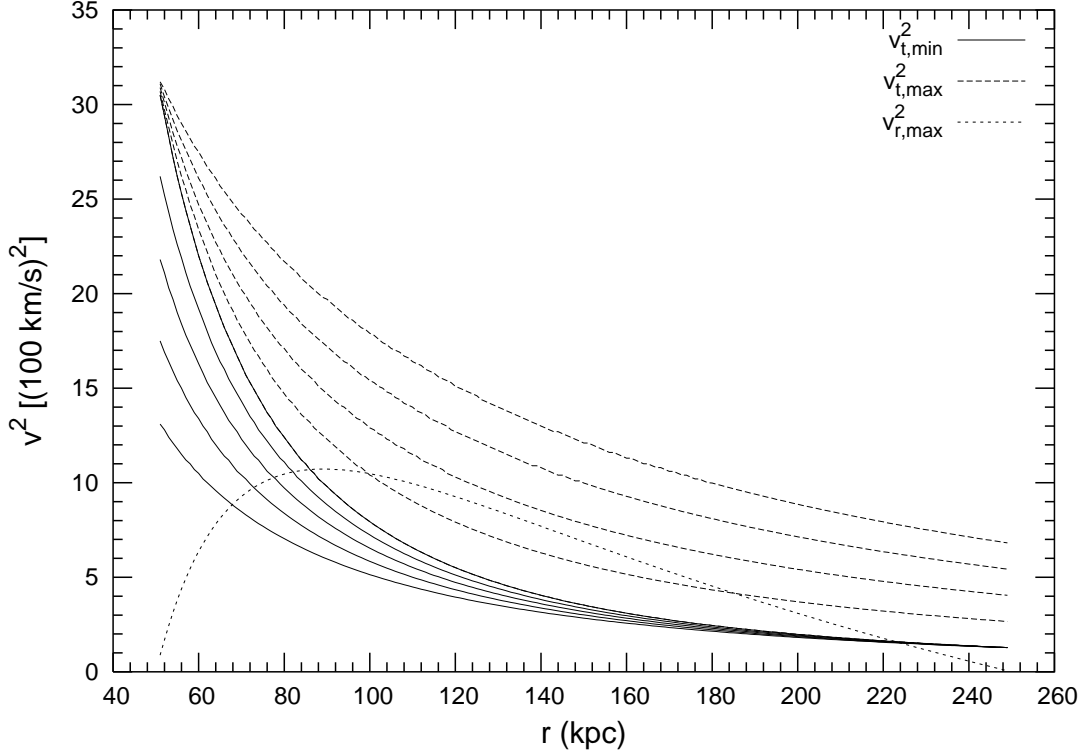}} 
\caption{Constraints on the radial and transverse velocities of dSphs 
given by equations
(\ref{vrmaxsq_eq}), (\ref{vtmaxsq_eq}) and (\ref{vtminsq_eq})
displayed as functions of $r$ for $\rmin=50\kpc$ and
$\rmax=250\kpc$, for the ``softened Keplerian'' potential of the
form $\phi(r)=-\kappa \left(1+(r/r_0)\right)^{-1}$ with
normalizations $\kappa\sim 40\times (100\kmps)^2$ and $r_0\sim
65\kpc$. The $\vtminsq(r,\vr(r))$ (solid curves) and
$\vtmaxsq(r,\vr(r))$ (dashed curves) are shown for five different
values of $v_r^2(r)$, namely, $v_r^2(r)=(0.2, 0.4, 0.6,
0.8,1)\times\vrmaxsq(r)$ (from top to bottom). The dotted curve
shows $\vrmaxsq(r)$. Similar constraints derived from a self-consistent 
model of the Galaxy were used for deriving the results quoted in the 
paper.} 
\label{fig:v2_limits_fig}
\end{figure}

The constraint equations (\ref{vrmaxsq_eq}), (\ref{vtmaxsq_eq}) and
(\ref{vtminsq_eq}) as functions of $r$ are displayed for
illustration in Figure \ref{fig:v2_limits_fig} for a simple
representative Galactic potential at large $r$ of the form
\begin{equation}
\phi(r)=-\kappa
\left(1+(r/r_0)\right)^{-1}\,,\label{phi_soft_kepler_pot}
\end{equation}
where $\kappa$ and $r_0$ are dimensionful constants. We have found
that this simple ``softened Keplerian'' potential with appropriately
chosen values of the normalization constants, $\kappa\sim 40\times
(100\kmps)^2$ and $r_0\sim 65\kpc$ roughly reproduces the potential of the 
Galaxy at large galactocentric distances $\gsim 
50\kpc$. Notice that $\vrmaxsq$ vanishes at $\rmin$ and $\rmax$ and is 
less than $\vtmaxsq$ for considerable distances from the end points, and 
it is only at intermediate distances that $\vrmaxsq$ becomes larger than 
$\vtminsq$ and $\vtmaxsq$ for large values of $\vr$. Thus it is clear that 
such kinematic restrictions on the population will yield one with a large 
value of the asymmetry parameter $j$ defined in equation (\ref{j_def1}). 
We should emphasize that the choice of the potential 
(\ref{phi_soft_kepler_pot}) here is merely 
to illustrate the general nature of the constraint equations. In the 
actual calculations and the results presented in this paper the potential 
is calculated self-consistently using a dynamical model of the Galaxy 
described in section \ref{sec:dmhalo_model_math}. 

The kinematic constraints derived above will immediately impose 
asymmetries on the phase space distribution function of the dSphs.  
Writing the primitive isotropic phase space distribution function of the 
dSphs as $F(\vt,\vr,r)$, the spatial number density distribution of the 
dSphs is given by 
\begin{equation}
n(r)= \int_{0}^{\vrmax}\int_{\vtmin}^{\vtmax}
F(\vt,\vr,r) \,\, 2\pi \,\, \vt \,\, d\vt \,\, d\vr\,,
\label{n_eqn}
\end{equation}
for $\rmin < r < \rmax$, and $n(r)=0$ otherwise. The radial number 
distribution $N(r)$ of the dSphs is then given by equation 
(\ref{N_r_dr_def}). 

Similarly the velocity skewness or asymmetry parameter $j$ defined in 
equation (\ref{j_def1}) is given by 
\begin{equation}
j =
\frac{\int_{\rmin}^{\rmax}\int_{0}^{\vrmax}\int_{\vtmin}^{\vtmax}
F(\vt,\vr,r)v^2 \,\, 2\pi \,\, \vt \,\, d\vt \,\, d\vr \,\, d^3r}
{\int_{\rmin}^{\rmax}\int_{0}^{\vrmax}\int_{\vtmin}^{\vtmax}
F(\vt,\vr,r)v_r^2 \,\, 2\pi \,\, \vt \,\, d\vt \,\, d\vr \,\,
d^3r}\,. \label{j_def2}
\end{equation}

Before we calculate these constrained distributions and the skewness 
parameter from a self-consistent dynamical model, we will discuss the 
importance of the skewness parameter $j$ in constraining the circular 
speed and in turn the Galactic potential at large distances. 
\section {Circular rotation speed at large Galactocentric distances
\label{sec:virial}}
The estimate of the skewness parameter $j$ from the observed data 
allows us to derive the circular rotation speeds $v_c$ in the
Galactic potential at distances spanned by the dSphs. Following
\citet{LF81} consider the identity
\begin{equation}
\half\frac{d^2|\mbox{\boldmath $r$}|^2}{dt^2} - {\mbox{\boldmath
$v$}}^2 = \mbox{\boldmath $r$}.\ddot{\mbox{\boldmath
$r$}}=\mbox{\boldmath $r$}. \mbox{\boldmath $\nabla$}\phi=-v_c^2\,,
\label{LF_identity}
\end{equation}
applicable to any particle moving in a gravitational potential
$\phi$. Here $\mbox{\boldmath $v$}=\dot{\mbox{\boldmath $r$}}$
denotes the velocity vector of a particle at the position
$\mbox{\boldmath $r$}$ with respect to the center, and $v_c(r)$ is
the circular velocity that would balance the radial component of
gravity at the radius $r$. Averaging equation (\ref{LF_identity})
over an ensemble of particles with position vectors $\mbox{\boldmath
$r_i$}$ and noting that
\begin{equation}
\frac{d}{dt}\sum_{i}\mbox{\boldmath $r_i$}^2=\frac{dI}{dt}=0\,
\label{mom_inertia_eqn}
\end{equation}
for a system of particles in virial equilibrium ($I$ being the
moment of inertia), one gets
\begin{equation}
\langle v^2\rangle = \langle v_c^2\rangle\,. \label{LF_theorem}
\end{equation}

Thus, by measuring $\vdispsq$ for an ensemble of objects at large
Galactocentric distances we may estimate the value of $\vcdispsq$ at
those distances. As emphasized by \citet{LL95}, the validity of this
theorem does not require that the potential be self-generated by the
ensemble of particles under consideration. Thus, in general, we can
write,
\begin{equation}
\vcdispsq = j \vrdispsq\,, \,\,\,{\rm or} \,\,
v_c\sim j^{1/2}\vrdispsq^{1/2}\,.
\label{vc_j_eqn}
\end{equation}

It is worth emphasizing that in this formulation of the virial theorem 
the kinetic energy gets equated to $\langle r \partial\phi/\partial r 
\rangle$ rather than to the potential energy as in the standard
formulation.  
 
The root mean square of the radial velocities, $\vrdisp$, can be 
estimated from a compilation of the radial velocities of various
astronomical objects given in Table 2 and 3 of WE99, for example.
This gives $\vrdisp\sim115\kmps$. The errors we expect in the
determination of this $\vrdisp$ are primarily systematic, even
though the smallness of the sample will also contribute to the
uncertainties. Referring to the extensive review by \citet{Mateo98}
we note that the typical accuracy with which the radial velocities
with respect to the solar system barycentre are quoted is $\pm 2\kmps$. 
Thus the errors are primarily due to the
uncertainty in the circular velocity of the local standard of rest.
Keeping in mind that the value of the circular rotation velocity at
solar circle quoted in literature is in the range 200 -- 220
$\kmps$, and that the component of this velocity in the direction of the 
satellite is to be subtracted from the observed radial velocity with 
respect to the solar system, there would be an additional uncertainty of 
$\sim 5\kmps$ on the average in the estimate of $\vrdisp$. 

The measurements of proper motions are less certain 
and provide estimates of 
the transverse velocities as seen from the solar system. Since the solar 
system is at a distance of $\sim 8.5\kpc$ from the Galactic centre we need 
to make use of both the observed components of the velocities to determine 
the radial velocity with respect to the Galactic centre. In so far as the 
distances of the satellites are much larger than 8.5 kpc, the 
transformations of the velocity components from the heliocentric to 
the Galactocentric system do not add additional uncertainties except in 
rare instances. Thus the determination of $\vrdisp$ is reasonably robust. 

On the other hand, the transverse velocities are to be estimated from the 
measurements of proper motions and are subject to greater uncertainties. 
This, coupled with the fact that there are only a small number of dSphs 
for which proper motion data are available, makes reliable observational 
determination of the anisotropy parameter $j$ from direct 
measurements of the radial and transverse velocities somewhat difficult. 
We shall instead estimate the parameter $j$ in section 
\ref{sec:j_estimation} below from the observed radial 
number distribution, $N(r)$, of the dSphs, which was introduced in section 
\ref{sec:spatial_df_sats}. 

To proceed further, we next describe the model we have adopted for
calculating the potential of the Galaxy in which the dSphs move. 
\section {Dynamical Model for the Dark Matter Halo of the Galaxy: The 
mathematical formalism
\label{sec:dmhalo_model_math}}
Both the visible matter and the dark matter contribute to the potential of 
the Galaxy, even though their relative contributions vary, with visible 
matter dominating at galactocentric distances below $\sim 5\kpc$ and the 
dark matter contribution increasing slowly with distance until beyond 
$\sim 20\kpc$ it is the dominant contributor. Whereas the density 
distribution of visible matter may be directly inferred from the 
astronomical observations, the density distribution of dark matter has to 
be deduced by somewhat more involved methods. The spatial distribution 
of the DM particles is dictated by their velocity distribution and the 
overall gravitational 
potential to which they also contribute depending on their density 
distribution. Thus one should ensure internal consistency between their 
velocity distribution and their density whose contribution to the 
potential when added to that of the visible matter should yield the 
overall potential. 

The overall gravitational potential of the Galaxy is well determined by 
the observed rotation curve of the Galaxy at least up to $\sim$ 15 kpc. 
The method using the dSphs discussed in the previous section will help us 
to determine the potential up to distances of $\sim$ 100 -- 200 kpc. In 
order to ensure the overall self-consistency we need a dynamical model, 
and to this end we start by making a convenient
ansatz regarding the functional form of the phase space distribution
of the particles constituting the DM halo. We then derive the
structure of the halo by solving the combined Poisson-Boltzmann
equation which involves the known gravitational potential of the visible 
matter of stars and gas of the Galaxy and the as yet undetermined 
potential of the DM halo. The solutions are a parametrized family of 
functions representing the potential due to the dark matter distribution. 
We may add to this the known contribution of the visible matter to get the 
total potential corresponding to different values of the parameters that 
characterize our DM halo model. By fitting the potential derived from the  
measurements of the rotation curve and the dSph dynamics, we can then 
determine the acceptable range of the parameters of our assumed DM halo 
model.   

We assume that the density distribution of the normal visible matter
is known and can be described adequately by a spheroidal bulge
superposed on an axisymmetric disc~\citep{CO81,BT87,KG89}. The
density distributions of these are given respectively by

\begin{equation}
\rho_s(r)=\rho_s(0)\left(1 +\frac{r^2}{a^2}\right)^{-3/2}\,,
\label{rho_sph}
\end{equation}

and

\begin{equation}
\rho_d(r)=\frac {\Sigma}{2h}e^{-(R-R_0)/R_d} \,\,\, e^{-|z|/h}\,,
\label{rho_disk}
\end{equation}

where $r=(R^2+z^2)^{1/2}\, $, $R$ being the Galactocentric distance
in the median plane of the disc and $z$ the distance normal to the
plane. The parameters take the values $\rho_s(0)=4.2 \times
10^{2}\msun\pc^{-3}\,\,$, $a=0.103\kpc\,$, $R_d=3.5\kpc\,$, and
$h=0.3\kpc\,$. Also $R_0=8.5\kpc\,$ is the solar Galactocentric
distance and $\Sigma\approx 36\msun\pc^{-2}$ is the column density
of the disk at the solar location. The expressions for the
gravitational potentials, $\phi_s$ and $\phi_d$, corresponding to
above forms of $\rho_s$ and $\rho_d$, are given in \citet{CO81} and
\citet{KG89}.

The true phase space DF that describes the DM halo of the Galaxy is not 
known. To make progress we choose a phase-space distribution function (DF) 
of the dark matter dictated by the following physical considerations: It 
should represent a collisionless system and should allow a parametrization 
in terms of the three main physical parameters of the halo, namely, the 
density and velocity 
dispersion of dark matter at some reference location within the halo, and 
the size (radius) of the halo. The truncated or the so-called ``lowered'' 
isothermal distribution (often called the ``King'' model) is a simple DF 
which has these features, and has been studied extensively 
\citep[see, e.g.,][p.~232]{BT87}. As such, in the present study we adopt 
this DF, which is given by 

\begin{equation}
f(x,v) \equiv f(\varepsilon)=\left\{ \begin{array}{ll} \rho_1
(2\pi\sigma^2)^{-3/2}\left(e^{\varepsilon/\sigma^2} -1\right) &
\mbox{for $\varepsilon > 0$}\,,\\
0 & \mbox{for $\varepsilon\leq 0$}\,,
\end{array}
\right. 
\label{king_df}
\end{equation}
with
\begin{equation}
\varepsilon\equiv\phi_0 - (\half v^2 + \phi)\,.
\label{relative_energy_def}
\end{equation}

Here $\phi=\phi_s + \phi_d + \phidm$ is the {\it total}
gravitational potential due to the visible and the dark matter
components of the Galaxy, and $\phi_0$, $\rho_1$ and $\sigma$ are
parameters to be determined by comparing the results of the model
calculations with observations. Notice that here $f$ is chosen as a
function of the conserved quantity, total energy (per unit mass),
$E=(\half v^2 + \phi)$, so that the collisionless Boltzmann equation
is stationary. The density of dark matter is given by
\begin{equation}
\rhodm=\int f \,\, d^3v\,, \label{rho_dm_eqn}
\end{equation}
and vanishes at the location $r=r_t$ where $\varepsilon=0$,
representing the outer edge of the halo. Note also that the parameter 
$\sigma$ (having the dimension of velocity) is {\it not} equal to the 
velocity dispersion of the DM particles, $\vdispdm$; the latter can be 
calculated for the DF given above and is a function of $r$, vanishing 
at $r=r_t$.   

The visible matter potentials and densities being already known, the
DM potential $\phidm$ is obtained by solving the Poisson equation
\begin{equation}
\nabla^2\phidm(R,z)=4\pi G \rhodm(R,z)\,. \label{phi_dm_eqn}
\end{equation}

Notice that $\rhodm$ appearing on the $r.h.s.$ of the equation
(\ref{phi_dm_eqn}) is a nonlinear functional of $\phi=\phi_s +
\phi_d + \phidm$ via equations (\ref{king_df}),
(\ref{relative_energy_def}) and (\ref{rho_dm_eqn}). Assuming 
axisymmetry, we numerically
solve the nonlinear equation (\ref{phi_dm_eqn}) through an iterative
procedure discussed earlier in \citet{CRB96}, which has been tested on 
special cases where analytical
solutions are available. The solutions thus obtained give the
values of $\phidm(R,z)$ for different chosen values of the
parameters $\rho_1, \sigma$ and $\phi_0$.

In our numerical calculations we take the DM density at the solar  
location, $\rhodmsun=\rhodm(R=R_0,0)$, the DM velocity dispersion at the 
solar location, $\vdispdmsun=\vdispdm(R=R_0,0)$, and the truncation radius 
$r_t$, as the three ``observable'' free parameters of the model, instead 
of the parameters $\rho_1$, $\sigma$ and $\phi_0$ appearing in equations 
(\ref{king_df}) and (\ref{relative_energy_def}). From this 3-parameter 
family of solutions we need to choose the one which fits all the known 
observations.  

\section {The Rotation Curve and the value of $\rhodmsun$ 
\label{sec:rotation_curve}}
One of the most useful probes of the potential of the Galaxy is the 
circular rotation curve (RC), $v_c(R)$, which is given by
\begin{equation}
v_c^2 (R) = R \frac{\partial\phi}{\partial R}(R,0) =
R\frac{\partial}{\partial R}\left[\phidm(R,0)+\phi_s(R,0)+
\phi_d(R,0)\right]\,. \label{v_c_def}
\end{equation}

\begin{figure}
\rotatebox{270}{\includegraphics[width=4in]{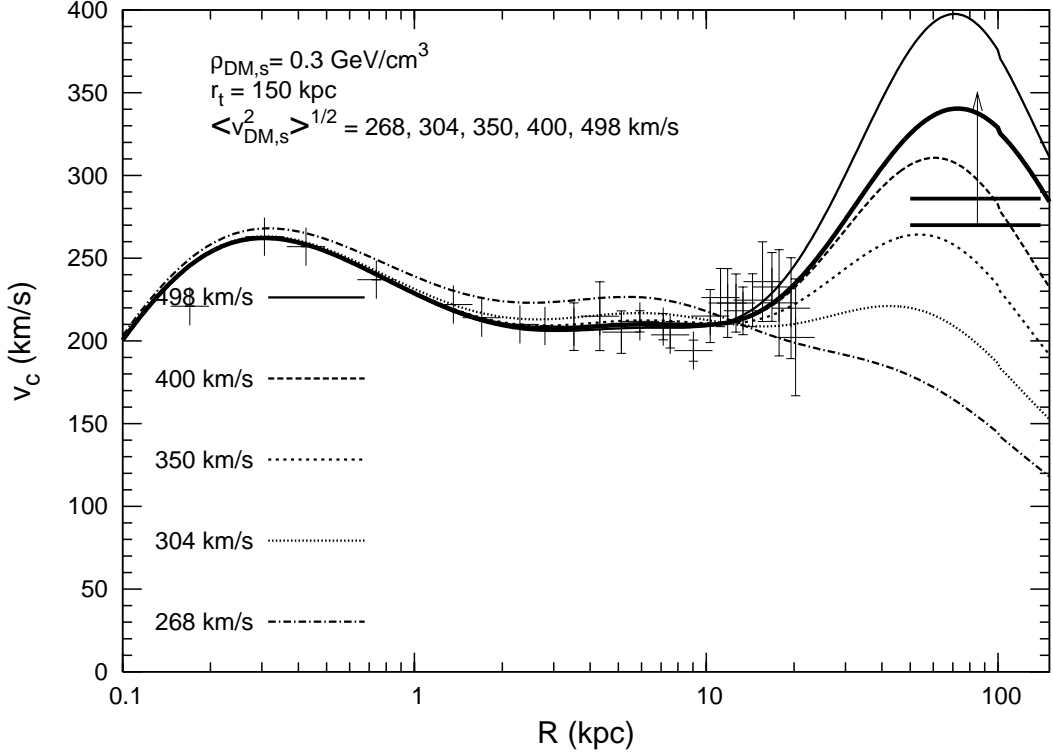}}
\caption{
A sample of theoretically calculated rotation curves of the Galaxy based 
on our self-consistent model, including the lower 
bounds derived from the study of dynamics of dwarf spheroidals. The dark 
matter is modeled as having a lowered (truncated) isothermal distribution. 
The curves shown are for dark matter density in the solar neighborhood,  
$\rhodmsun\approx 0.3\gev\cm^{-3}$, truncation radius, $r_t=150\kpc$,  
and various values of $\vdispdmsun$, the velocity dispersion of the dark 
matter particles at the solar neighborhood, as indicated. The solid 
curve corresponds to the maximum possible value of $\vdispdmsun$ 
consistent with the chosen values of $\rhodmsun$ and $r_t$. The doubly 
thick solid curve represents our ``conservative most likely'' (CML) model 
(see text) with DM parameter values $\rhodmsun\approx 0.3\gev\cm^{-3}$, 
$\vdispdmsun\approx 400\kmps$ and $r_t=200\kpc$.    
The observational data (crosses and vertical lines with error bars) are 
from \citet{HS97} for the case 
$R_0=8\kpc$ and $v_c(R=R_0)=200\kmps$. The two solid horizontal lines  
represent two different estimates of the expected lower limits on 
$v_c$ in the region of $R\sim 100\kpc$ obtained in this paper from the 
dynamics of the dwarf spheroidals. The small glitches in the curves at 
$R\simeq100\kpc$ are artifacts of numerical calculation and are due 
to increased grid spacing used for distances $r\geq 100\kpc$ in order to  
reduce the total computation time.} 
\label{fig:rc_fig_mod} 
\end{figure}

We have run a series of models of the Galaxy spanning a wide range 
of values of the parameters $\rhodmsun$, $\vdispdmsun$ and $r_t$ 
distributed about their values estimated from earlier studies. 
Figure \ref{fig:rc_fig_mod} shows a sample of the RC's of the
Galaxy calculated from our model~\footnote{In Appendix 
\ref{apndx:model_rotcurves} we present several more of these RCs  
calculated for values of the relevant parameters of the 
halo neighboring those used in Fig.\ref{fig:rc_fig_mod}.} together with 
the available
data for $R\lsim20\kpc$ as summarized in \citet{HS97}. In the same Figure 
we have also indicated two different estimates of the expected lower 
limits on $v_c$ in the region $R\sim 100\kpc$ obtained in this paper from the 
dynamics of the dwarf-spheroidals; derivation of these estimates is 
discussed in section \ref{sec:j_estimation} below. 

In the very central regions ($R\lsim 1\kpc$) of the Galaxy the density of 
normal matter is so high that the RC is essentially determined by it and 
is sensibly independent of the parameters of the DM. In the region $1\kpc 
< R \lsim 10\kpc$ the contribution of normal matter progressively 
decreases as the contribution of DM increases, maintaining a nearly flat 
RC. Based on the results of our extensive calculations of 
approximately more than two 
hundred model RCs for a wide range of the relevant parameters (only a 
selected few of which are shown in Fig.~\ref{fig:rc_fig_mod}, 
Fig.~\ref{fig:appendix_rc_fig_0.2} and 
Fig.~\ref{fig:appendix_rc_fig_0.4}), we can 
conclude that, as long as we restrict ourselves to the available data below 
galactocentric distances of $\sim 20\kpc$, a range of $\rhodmsun$ values 
from $\sim 0.25\gev/\cm^3$ to $\sim 0.4\gev/\cm^3$ with suitably  
chosen values of the other two parameters, $r_t$ and $\vdispdmsun$, can 
yield acceptable fits to the data. To be specific, we use the value     
\begin{equation}
\rhodmsun\approx 0.3\gev \cm^{-3}\,
\label{rhodmsun_value}
\end{equation}
for the DM density in the solar neighborhood in our 
numerical calculations discussed below; the results derived in this paper 
do 
not change by any significant amount for other values of this parameter 
within the approximate range given above. Some theoretical rotation 
curves for other 
values of $\rhodmsun$ are given in Appendix \ref{apndx:model_rotcurves}.     
 
Sensitivity of the rotation curves to the two other parameters of 
our DM halo model, namely, $r_t$ and $\vdispdmsun$, commences at 
distances $R > 10\kpc$, with strong sensitivity to these parameters 
at $R > 20\kpc$ where direct measurements of the rotation curve 
are absent. The estimates of the rotation speeds derived 
from the dynamics of dSphs, therefore, play an important role in 
constraining the values of these two parameters, which we shall proceed to 
discuss in the next section. 

Based on the nature of the theoretical rotation curves displayed in 
Fig.~\ref{fig:rc_fig_mod}, Fig.~\ref{fig:appendix_rc_fig_0.2} and 
Fig.~\ref{fig:appendix_rc_fig_0.4}, 
for example, together with the existing rotation curve data, we shall 
restrict our attention, for the subsequent 
discussions in the paper, to the values of the 
parameters $r_t$ and $\vdispdmsun$ satisfying  
\begin{equation}
r_t\ge 100\kpc\,,
\label{r_t_restriction} 
\end{equation}
and 
\begin{equation}
\vdispdmsun \ge 300\kmps\,. 
\label{vdispdmsun_restriction}
\end{equation}
For suitable choices of values of the parameters in these ranges, the 
theoretical rotation curves fit the observed data 
satisfactorily. Some theoretical rotation curves for parameter 
values outside the values and ranges indicated by equations 
(\ref{rhodmsun_value}), (\ref{r_t_restriction}) and  
(\ref{vdispdmsun_restriction}) are also shown  
in Fig.~\ref{fig:rc_fig_mod},  
Fig.~\ref{fig:appendix_rc_fig_0.2} and Fig.~\ref{fig:appendix_rc_fig_0.4}.

\begin{figure}
\rotatebox{270}{\includegraphics[width=4in]{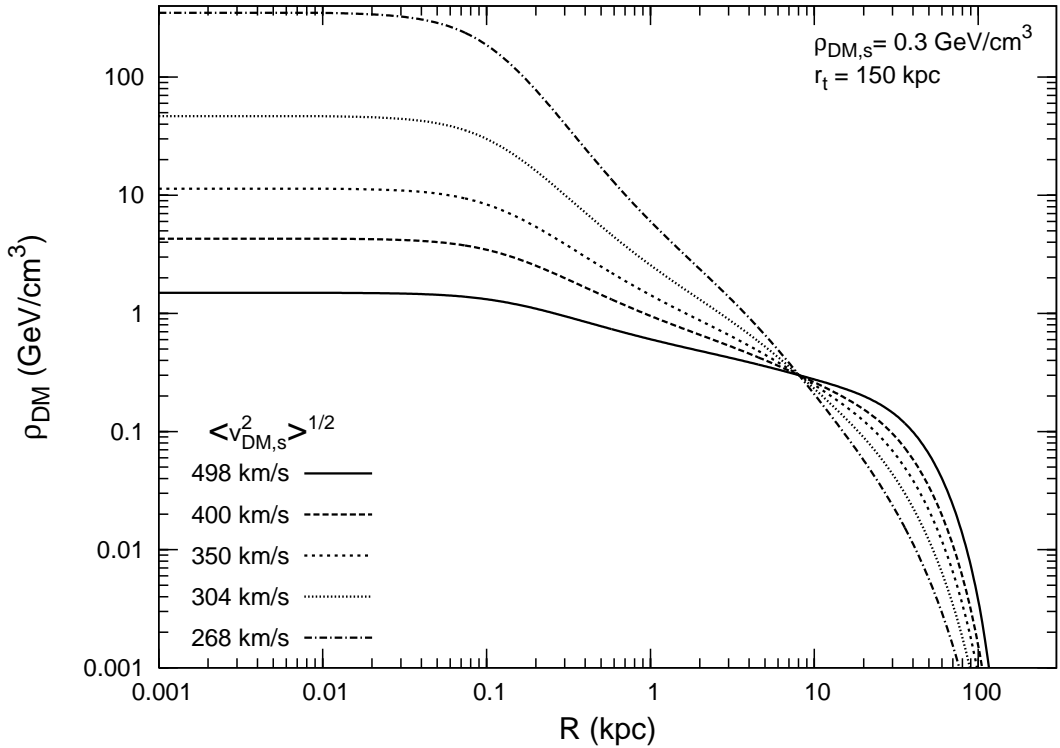}}
\caption{
The dark matter density profiles in the plane of the Galaxy for the 
same set of model parameters corresponding to the rotation 
curves of Fig.~\ref{fig:rc_fig_mod}. 
}
\label{fig:rhodm_fig} 
\end{figure}

The DM density profiles in the plane of the Galaxy for the same set of 
our self-consistent models corresponding to the rotation curves of 
Fig.~\ref{fig:rc_fig_mod} are shown in Figure \ref{fig:rhodm_fig}.  

\section {Constraining the parameters of the dark matter halo using the 
dynamics of dwarf spheroidals
\label{sec:constraints_from_dsphs}} 
As discussed in Section \ref{sec:virial} the rotation speeds at large 
galactocentric distances can be estimated from the dynamics of the dSphs. 
In sections \ref{sec:spatial_df_sats} and 
\ref{sec:truncated_df_dsph} it was shown that the dSphs will  
have a skewness 
in their velocity distribution quantified in terms of the parameter $j$, 
which is a 
sensitive function of the parameter $\rmin$, the location of the dip in 
the radial number distribution of the satellites of the Galaxy. We can 
determine the value of $\rmin$, and hence, the parameter $j$, from the 
observed radial distribution of the dSphs in the following way. 

Referring to Fig.~\ref{fig:hist_all_TF}, let $N_i^{\rm obs}$ be the 
observed number of dSphs in the $i$-th, 10 kpc-size radial distance bin,  
with $i=1,2,\ldots$ denoting the radial distance bins (0--10 kpc), 
(10--20 kpc), etc., respectively. We calculate the theoretical 
expectation for the number 
of objects in each of these bins, $N_i^{\rm th}$, from the radial number 
distribution of the 
satellites, $N(r)$, calculated in a number of theoretical models 
with various possible values of $\rmin$ and a large enough value 
of $\rmax\approx 500\kpc$. Our calculations discussed below have 
negligible dependence on the exact value of $\rmax$ as long as it is 
large enough to accommodate the most distant satellites. 

The radial number distribution $N(r)$ can be calculated using equations 
(\ref{N_r_dr_def}) and (\ref{n_eqn}). To do this, we need to provide  
the primitive (undistorted) phase space distribution function, 
$F(\vt,\vr,r)$, of the satellites, which we take to be of isothermal 
form, namely, 
\begin{equation}
F(\vt,\vr,r)\propto 
\exp\left\{-3\left(\frac{1}{2}(v_r^2+v_t^2)+\phi(r)\right)/\sigma^2_s\right\}\,,
\label{F_sat}
\end{equation} 
where $\sigma_s$ is the primitive velocity dispersion of the satellites, 
and $\phi(r)$ is the gravitational potential in which the satellites 
move~\footnote{
The DF (\ref{F_sat}) has an isotropic velocity 
distribution. We have investigated in Appendix \ref{apndx:j_aniso_DF} the 
effects of a possible anisotropic initial DF favoring radial 
velocities over the transverse component. As discussed there, our  
results do not change significantly compared to those obtained using the 
isotropic DF (\ref{F_sat}).}.  
The potential can be taken to be spherically symmetric without much 
loss of accuracy since our theoretically calculated self-consistent 
potentials become progressively spherically symmetric at the large 
galactocentric distances where the dSphs lie. The kinematic velocity 
limits appearing in 
equation (\ref{n_eqn}) are obtained by solving the constraint equations  
(\ref{vrmaxsq_eq}), (\ref{vtmaxsq_eq}) and (\ref{vtminsq_eq}) for a given 
self-consistent potential $\phi(r)$ and given values of $\rmin$ and 
$\rmax$. 

Having already fixed the DM halo model parameter $\rhodmsun$,  
we can numerically calculate the theoretical quantities 
$N_i^{\rm th}$ for plausible values of the other relevant parameters, 
namely, $r_t$, $\vdispdmsun$, $\sigma_s$ and $\rmin$, and determine the 
most likely values of these parameters by performing a simple likelihood 
analysis as follows: Assuming Poissonian probability distribution, 
\begin{equation}
P_i\left(N_i^{\rm obs}, N_i^{\rm th}\right)=e^{-N_i^{\rm 
th}}\left(N_i^{\rm th}\right)^{N_i^{\rm obs}}
\frac{1}{N_i^{\rm obs}!}\,,
\label{poisson_prob}
\end{equation}
for the occurrence of $N_i^{\rm obs}$ number of dSphs in the $i$-th radial 
bin when the expectation is $N_i^{\rm th}$, we calculate the  
likelihood function ${\mathcal L}$ defined as 
\begin{equation}
{\mathcal L}\equiv\frac{\prod_{i=1}^{i_{\rm max}} P_i\left(N_i^{\rm obs}, 
N_i^{\rm 
th}\right)}{\prod_{i=1}^{i_{\rm max}}P_i\left(N_i^{\rm obs}, N_{i,{\rm 
B}}^{\rm th}\right)}
=\prod_{i=1}^{i_{\rm max}} e^{-\left(N_i^{\rm th}-N_{i,{\rm B}}^{\rm 
th}\right)}
\left(\frac{N_i^{\rm th}}{N_{i,{\rm B}}^{\rm th}}\right)^{N_i^{\rm 
obs}}\,,
\label{lf_ratio}
\end{equation}  
where $i_{\rm max}$ denotes the maximum number of radial bins and  
$N_{i,{\rm B}}^{\rm th}$ is the mean number expected in the $i$-th bin 
calculated in a base model with fixed values of the parameters $r_t$,   
$\vdispdmsun$, $\sigma_s$ and $\rmin$. 
 
Note that, depending on the value of $\rmin$ for the model under 
consideration, there will be bins with $r<\rmin$ for which $N_{i,{\rm 
B}}^{\rm th}$ or both $N_{i,{\rm B}}^{\rm th}$ and $N_i^{\rm th}$ may be 
zero, in which cases ${\mathcal L}$ defined in equation (\ref{lf_ratio}) 
diverges or is ill-defined. This, of course, is an artifact, due to 
the sharp cut-off in the number distribution of the satellites at $\rmin$. 
A more mathematically rigorous procedure would be to impose, for 
example, an exponential cutoff of the number distribution of the 
satellites below $\rmin$, thus giving a small but finite number for the 
values of $N_{i,{\rm B}}^{\rm th}$ and $N_i^{\rm th}$ below $\rmin$. 
Another way to avoid the problem is to use the simple regularization 
procedure of adding a small constant $\epsilon$ to $N_i^{\rm th}$ in the  
numerical calculations of ${\mathcal L}$, which is the procedure we have 
followed in our calculations of ${\mathcal L}$. We have checked that the 
resulting numerical values 
of $\mathcal L$ have negligible sensitivity to the assumed value of 
$\epsilon$ in the range $\epsilon\sim 10^{-6}$ -- $10^{-4}$. Accordingly, 
we take $\epsilon=10^{-6}$ in our numerical calculations.  

\begin{figure}
\rotatebox{270}{\includegraphics[width=4in]{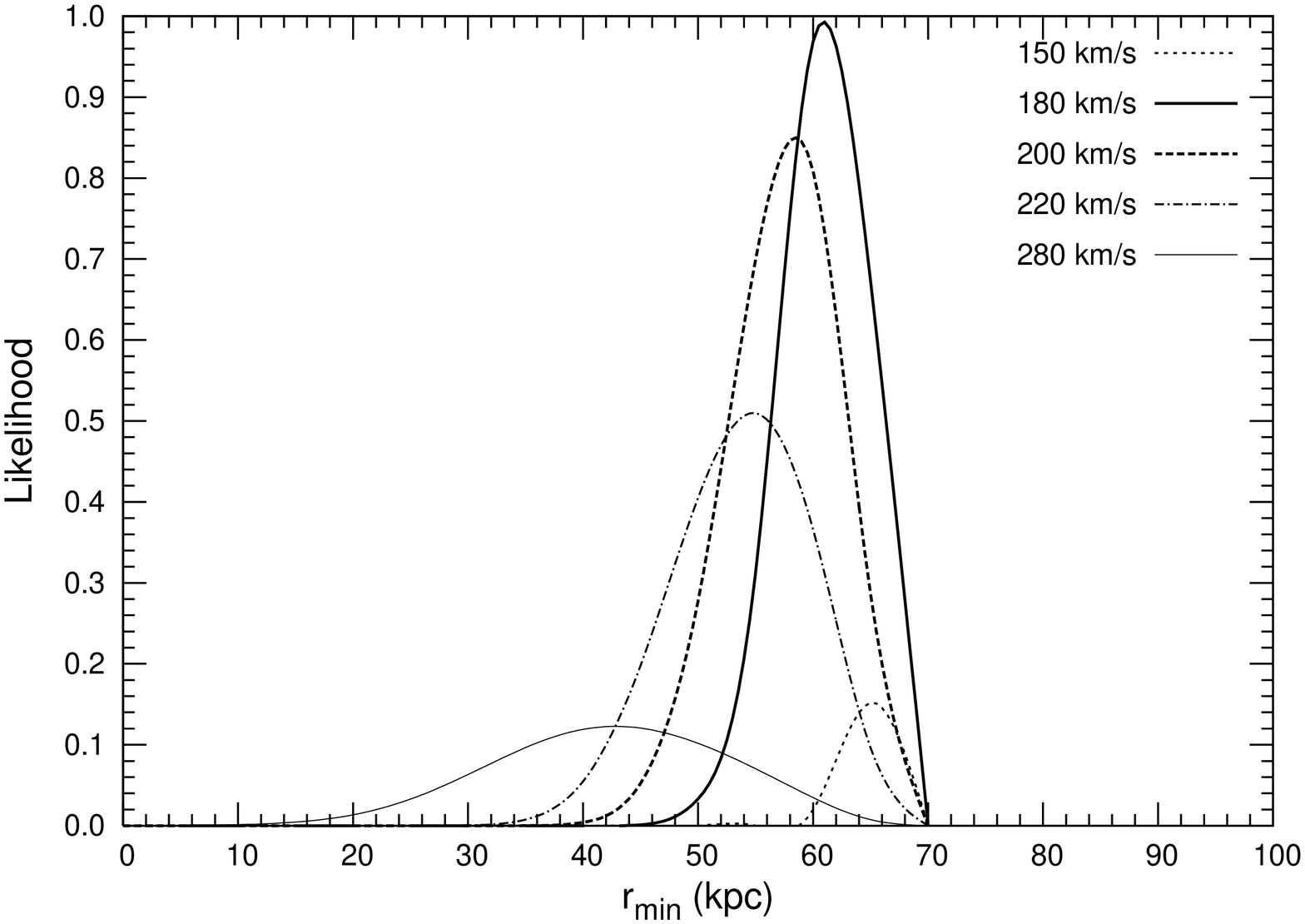}} 
\caption{The likelihood $\mathcal L$ as a function of the parameter 
$\rmin$ for various different values of $\sigma_s$ as indicated, for 
fixed values of $r_t=200\kpc$, $\rhodmsun=0.3\gev/\cm^3$ and 
$\vdispdmsun=400\kmps$. The base model has the parameter values 
$r_t=200\kpc$, $\rhodmsun=0.3\gev/\cm^3$, $\vdispdmsun=400\kmps$,  
$\sigma_s=180\kmps$ and $\rmin=61\kpc$.}
\label{fig:lfratio_sigmadsph_dm200_400}
\end{figure}

\subsection {Likelihood of model parameters  
\label{subsec:model_likelihoods}}
\begin{figure}
\rotatebox{270}{\includegraphics[width=4in]{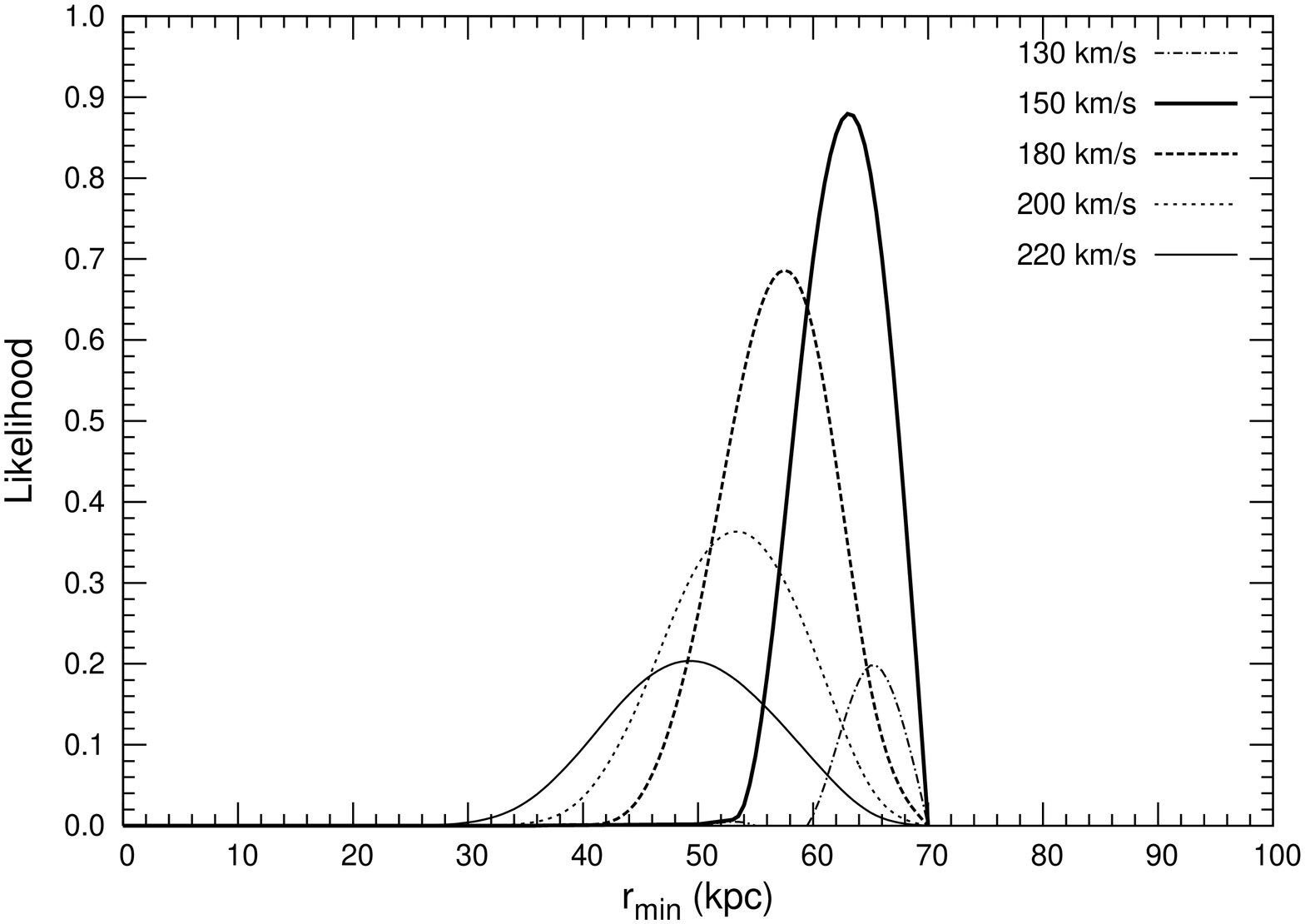}} 
\caption{The likelihood $\mathcal L$ as a function of the parameter 
$\rmin$ for various different values of $\sigma_s$ as indicated, for 
fixed values of $r_t=150\kpc$, $\rhodmsun=0.3\gev/\cm^3$ and 
$\vdispdmsun=400\kmps$. The base model has the parameter values 
$r_t=200\kpc$, $\rhodmsun=0.3\gev/\cm^3$, $\vdispdmsun=400\kmps$,  
$\sigma_s=180\kmps$ and $\rmin=61\kpc$.}
\label{fig:lfratio_sigmadsph_dm150_400}
\end{figure}

We have calculated the likelihood function $\mathcal L$ for a wide range 
of plausible values of the relevant parameters. 
Of the parameters $r_t$, $\vdispdmsun$, $\sigma_s$ and $\rmin$, the most 
sensitive dependence of $\mathcal L$ is on $\rmin$. In Figures  
\ref{fig:lfratio_sigmadsph_dm200_400}  and 
\ref{fig:lfratio_sigmadsph_dm150_400} we display the result of our 
calculation of $\mathcal L$ as a function of the parameter 
$\rmin$ for various different values of $\sigma_s$ and for two fixed sets 
of values of the DM model parameters, namely, $\rhodmsun=0.3\gev/\cm^3$,  
$\vdispdmsun=400\kmps$ and $r_t=200\kpc$ 
(Fig.~\ref{fig:lfratio_sigmadsph_dm200_400}), and 
$\rhodmsun=0.3\gev/\cm^3$, $\vdispdmsun=400\kmps$ and $r_t=150\kpc$ 
(Fig.~\ref{fig:lfratio_sigmadsph_dm150_400}). 

\begin{figure}
\rotatebox{270}{\includegraphics[width=4in]{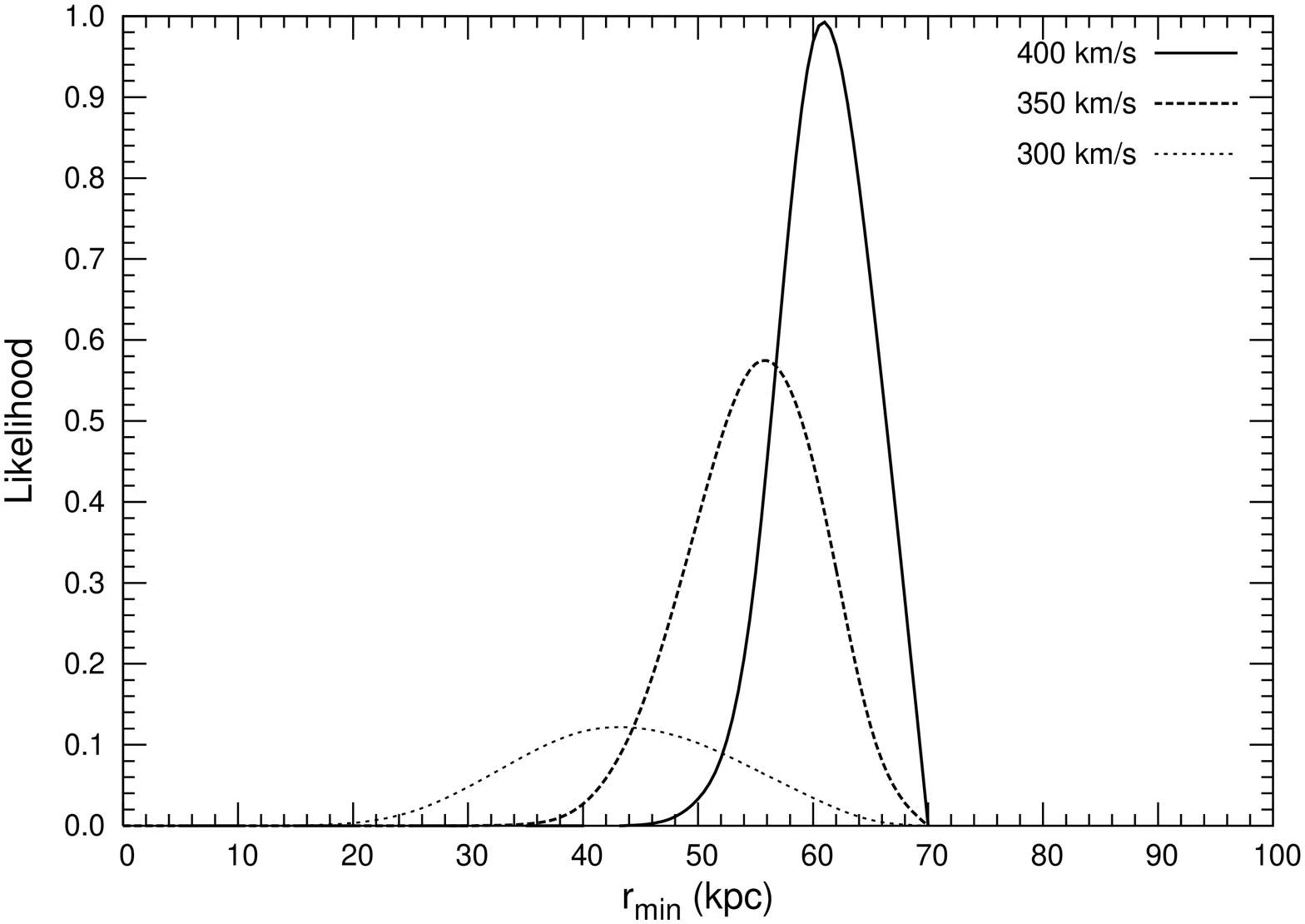}} 
\caption{The likelihood $\mathcal L$ as a function of the parameter
$\rmin$ for various different values of 
$\vdispdmsun$ as indicated, for
fixed values of $r_t=200\kpc$, $\rhodmsun=0.3\gev/\cm^3$ and
$\sigma_s=180\kmps$. The base model has the parameter values
$r_t=200\kpc$, $\rhodmsun=0.3\gev/\cm^3$, $\vdispdmsun=400\kmps$,
$\sigma_s=180\kmps$ and $\rmin=61\kpc$.}
\label{fig:lfratio_dmdisp_rt200}
\end{figure}

The dependence of the likelihood of different 
models on the value of the parameter $\vdispdmsun$ is illustrated in 
Figure \ref{fig:lfratio_dmdisp_rt200} where we display $\mathcal L$ as a 
function $\rmin$ 
for fixed values of $r_t=200\kpc$ and $\sigma_s=180\kmps$ but for 
different values of the DM parameter $\vdispdmsun$. 

The {\it base model} used in the likelihood curves shown in 
Figures \ref{fig:lfratio_sigmadsph_dm200_400}, 
\ref{fig:lfratio_sigmadsph_dm150_400} and \ref{fig:lfratio_dmdisp_rt200} 
has been chosen to have the parameter values 
\begin{equation}  
r_t=200\kpc\,,\, \rhodmsun=0.3\gev/\cm^3\,,\, \vdispdmsun=400\kmps\,, 
\label{base_model_DM_params}
\end{equation}  
\nopagebreak 
\begin{equation}
\sigma_s=180\kmps\,, {\rm and}\, \, \rmin=61\kpc\,.
\label{base_model_ds_params}
\end{equation} 

\begin{figure}
\rotatebox{270}{\includegraphics[width=4in]{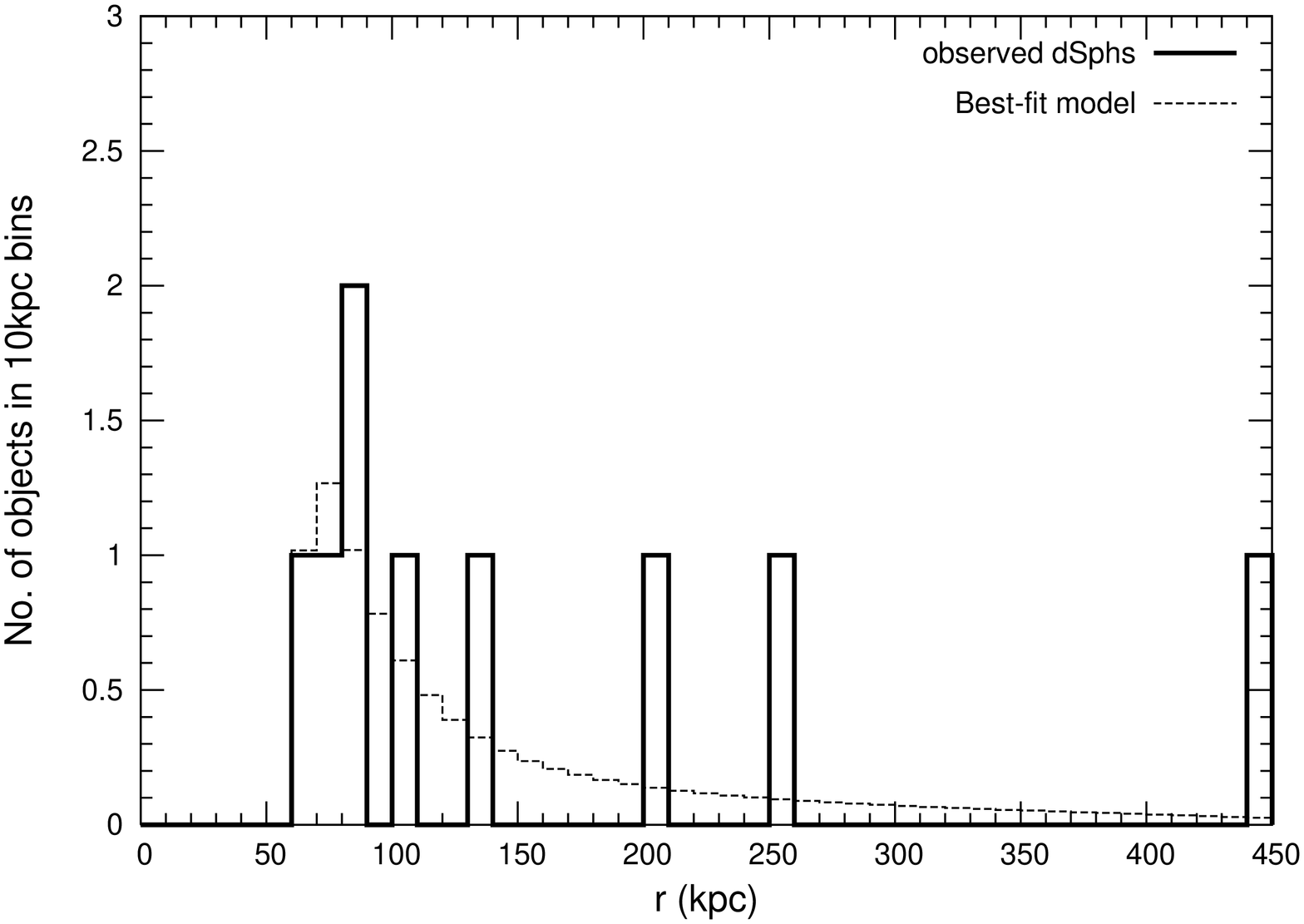}} 
\caption{Distribution of the number of dwarf-spheroidals as a function of 
the distance from the Galactic centre. The prediction of our 
conservative most likely model, which corresponds to 
$\vdispdmsun=400\kmps$, $\sigma_s=180\kmps$ 
and $\rmin=61\kpc$, with $\rhodmsun=0.3\gev/\cm^3$ and $r_t=200\kpc$, is 
also shown (see also Fig.~\ref{fig:hist_all_TF}).} 
\label{fig:hist_dsph_CML}
\end{figure}

The reason for choosing this set of parameter values for the base model is 
clear from a comparison of Figures 
\ref{fig:lfratio_sigmadsph_dm200_400}, 
\ref{fig:lfratio_sigmadsph_dm150_400} and 
\ref{fig:lfratio_dmdisp_rt200}, which  
shows that this model has the highest likelihood within the 
ranges of values of the relevant parameters considered in these Figures. 
We have also calculated the likelihood $\mathcal L$ for parameter values 
beyond their ranges displayed in Figures
\ref{fig:lfratio_sigmadsph_dm200_400},
\ref{fig:lfratio_sigmadsph_dm150_400} and
\ref{fig:lfratio_dmdisp_rt200}. The 
likelihood can be higher than that for the base model chosen above if 
we allow $\vdispdmsun > 400\kmps$. For example, with 
respect to the above chosen base model, and with other parameters being 
equal, we get $\mathcal L\approx 1.04$ 
for a model with $\vdispdmsun\approx 450\kmps$ and $\sigma_s\approx 
220\kmps$. We have not done a fine-grained scanning of the entire 
parameter space spanned by  
all the relevant parameters for $\vdispdmsun > 400\kmps$, but our 
present calculations indicate that the model with the absolute highest 
likelihood, i.e., the ``best-fit'' model seems to correspond to a value 
of $\vdispdmsun$ close to $\sim 500\kmps$ with $\sigma_s\sim 290\kmps$.  
However, to be on the 
conservative side as far as the most likely value of the important 
parameter $\vdispdmsun$ is concerned, we shall refer to the above chosen 
base model with parameters given by
equations (\ref{base_model_DM_params}) and (\ref{base_model_ds_params}) as  
our ``conservative most likely'' (CML) model. The radial distribution of 
the dSphs calculated using equations (\ref{N_r_dr_def}) and (\ref{n_eqn}) 
for the parameter values corresponding to the above CML model is 
shown in Figure \ref{fig:hist_dsph_CML}.  

From Figures \ref{fig:lfratio_sigmadsph_dm200_400},
\ref{fig:lfratio_sigmadsph_dm150_400} and \ref{fig:lfratio_dmdisp_rt200}, 
we find that the relevant parameter values {\it outside} the ranges
\begin{equation}  
\vdispdmsun \gsim 350\kmps\,,\, \, 150 \lsim r_t \lsim 200\kpc\,,
\label{fifty_percent_likelihood_DM_intervals}
\end{equation}
\nopagebreak  
\begin{equation}
52 \lsim \rmin \lsim 66\kpc\,,\, \, 150 \lsim\sigma_s \lsim  
220\kmps\,,\, 
\label{fifty_percent_likelihood_ds_intervals}
\end{equation}
have {\it less than 50\%} likelihood of explaining the observed radial 
distribution of the dSphs. Demanding higher likelihood 
narrows down these ranges further. Note that for the conservative 
most likely model the 
likelihood that $\rmin < 50\kpc$ is very small, less than 4\%.   

\section {Estimating the velocity anisotropy parameter $j$ and 
limits on the rotation speed at large galactocentric distances  
\label{sec:j_estimation}}
\begin{figure}
\rotatebox{270}{\includegraphics[width=4in]{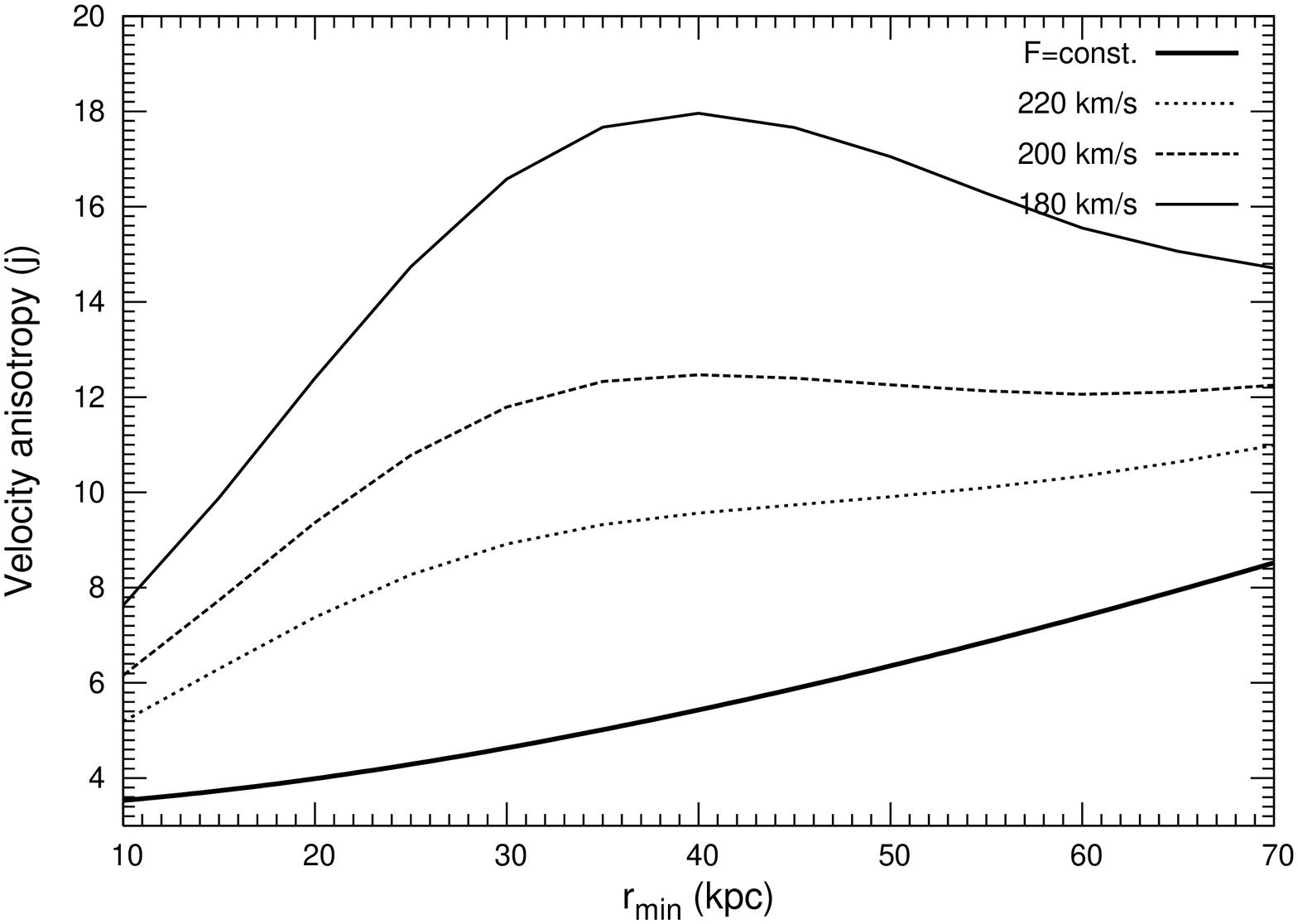}} 
\caption{The velocity anisotropy parameter $j$ as a function of $\rmin$ 
for various different values of $\sigma_s$ as indicated. The DM halo 
parameters have been fixed at $\rhodmsun=0.3\gev/\cm^3$, 
$\vdispdmsun=400\kmps$ and $r_t=200\kpc$.} 
\label{fig:j_fig_dm200_400}
\end{figure}

\begin{figure}
\rotatebox{270}{\includegraphics[width=4in]{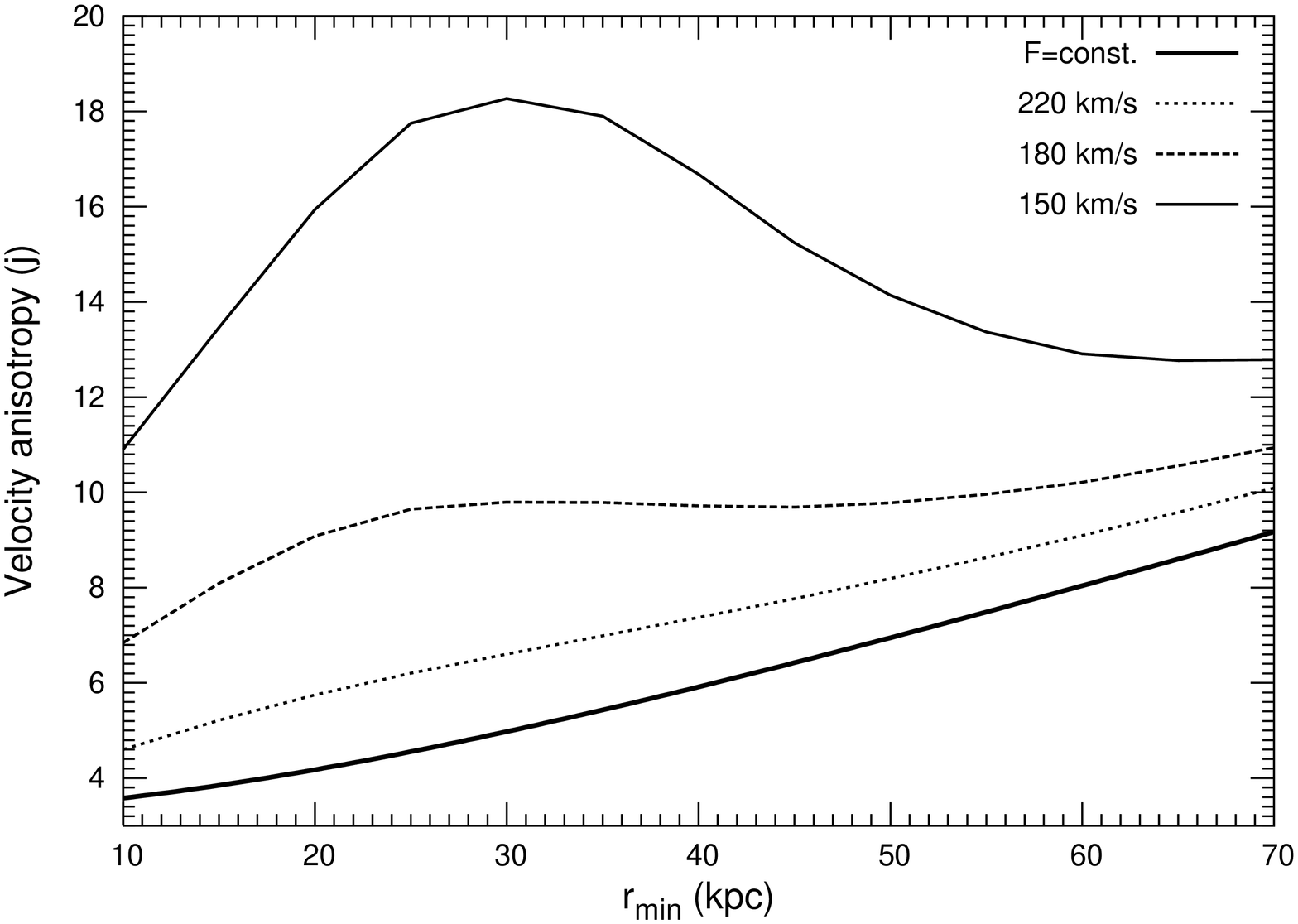}} 
\caption{Same as Fig.~\ref{fig:j_fig_dm200_400} but for 
$\vdispdmsun=350\kmps$ and $r_t=150\kpc$.} 
\label{fig:j_fig_dm150_350}
\end{figure}

We now have at hand all the parameters needed to theoretically estimate 
the value of $j$ using equation (\ref{j_def2}). To explicitly show the 
dependence of $j$ on $\rmin$ and $\sigma_s$, we have calculated $j$ as a 
function of $\rmin$ for various different values of $\sigma_s$, including 
very large values of $\sigma_s$ corresponding to a constant DF for the 
satellites. The results are shown in Figures \ref{fig:j_fig_dm200_400} 
and \ref{fig:j_fig_dm150_350} for $\vdispdmsun=400\kmps$, 
$r_t=200\kpc$ (Fig.~\ref{fig:j_fig_dm200_400}) 
and $\vdispdmsun=350\kmps$, $r_t=150\kpc$ 
(Fig.~\ref{fig:j_fig_dm150_350}). 

Note that, for a given potential $\phi(r)$ and a given value of $\rmin$, 
the lowest value of $j$ obtains, as expected, in the case 
$F=\,$ constant (i.e., $\sigma_s \to \infty$ in equation (\ref{F_sat})), 
because then all orbits in the interval $\rmin$ to $\rmax$ will be equally 
probable. Smaller (finite) values of $\sigma_s$ correspond to favoring 
orbits with mean apogalacticons smaller than $(\rmax - \rmin)/2$, making 
the orbits more circular and thus yielding larger values of $j$. 

Considering the dependence of $j$ on the DM halo model parameters 
$\vdispdmsun$  
and $r_t$, we find that $j$ depends most sensitively on $\vdispdmsun$ and 
weakly on $r_t$. In general, for a given value of $\vdispdmsun$, the value 
of $j$ is smaller for a smaller value of $r_t$, but only marginally so for 
values of $r_t$ in the range 100 -- 200 kpc. 

From Figure \ref{fig:j_fig_dm200_400} we find that for our CML model 
with the relevant parameter 
values given by equations (\ref{base_model_DM_params}) and 
(\ref{base_model_ds_params}), we get $j_{\rm CML}\approx 
15.4$. The observational data on the dSphs given in WE99 yield 
$\vrdisp=116\kmps$. Taking this to imply (conservatively) $\vrdisp\ge 
100\kmps$ one finds, using equation (\ref{vc_j_eqn}), 
\begin{equation}
\vcdisp_{\rm CML}=j^{1/2}_{\rm CML}\vrdisp\gsim 392\kmps\,.
\label{vc_CML}
\end{equation}
This value is consistent to within about 15\% with the peak value 
of the rotation curve for 
the DM model parameters corresponding to the CML model (indicated by the 
doubly thick solid curve in Figure \ref{fig:rc_fig_mod}). 

To be conservative on the side of the asymmetry parameter $j$, if 
we take the parameter values $\rmin\ge 50\kpc$, 
$\sigma_s \le 220\kmps$, $\vdispdmsun \ge 350\kmps$ and $r_t \ge 150\kpc$, 
then we get 
$j_{\rm conservative}=8.2$, yielding  
\begin{equation}
\vcdisp_{\rm conservative} \ge 286\kmps
\label{vc_conservative1}
\end{equation} 
as a conservative estimate of the rotation speed 
at $r\sim$~100--200~kpc, the typical distances of the dwarf
spheroidals.

Even more conservatively, taking $\sigma_s \le 280\kmps$ with $\rmin\ge 
50\kpc$, $\vdispdmsun \ge 350\kmps$ and $r_t \ge 150\kpc$, we get $j=7.1$, 
which yields   
\begin{equation}
\vcdisp_{\rm most\,\, conservative} \ge 266\kmps\,.
\label{vc_conservative2}
\end{equation} 
We have indicated the conservative lower limits (\ref{vc_conservative1}) 
and (\ref{vc_conservative2}) in Figures \ref{fig:rc_fig_mod},
\ref{fig:appendix_rc_fig_0.2} and \ref{fig:appendix_rc_fig_0.4} 
by the solid 
horizontal lines. Note that these lower limits on the rotation speeds are 
obtained entirely from the dynamics of the dSphs, and are clearly 
consistent with the rotation curves shown in Figures \ref{fig:rc_fig_mod}, 
\ref{fig:appendix_rc_fig_0.2} and 
\ref{fig:appendix_rc_fig_0.4}  for 
$\vdispdmsun > 350\kmps$ and $r_t > 150\kmps$. 

In our analysis thus far in the estimation of $j$ we have assumed the DF 
of the dSphs to be isotropic at the epoch of their formation. In Appendix 
\ref{apndx:j_aniso_DF} we investigate the effects of making the initial DF 
favor radial velocities over transverse velocities. This analysis 
indicates that once the DF is modified by removing the subset of orbits 
with small perigalacticons, the resultant distribution function rapidly 
starts favoring transverse velocities. An intuitive understanding of the 
generation of anisotropies favoring transverse velocities as a result of 
imposing a minimum value on the perigalacticon radius may be gained 
through a kinematic analysis of elliptical orbits; this is presented in 
Appendix \ref{apndx:j_ellipse}. 

From the foregoing analysis, supplemented by the discussions in the 
appendices A, B and C, we find that we are able to fit the available data 
on the rotation curve of the Galaxy, the radial distribution of the dwarf 
spheroidals and the value of the circular rotation speed at $\sim 100\kpc$ 
estimated from the analysis of the data on dSphs, with a self-consistent 
model of the Galactic dark matter halo in which the phase space 
distribution function of the DM particles is described by the King (i.e., 
the truncated isothermal) model, with the following values of 
the model parameters: 
\begin{equation}
\begin{array}{rll}
\rhodmsun & \approx & 0.3\gev\cm^{-3}\,, \\
\vdispdmsun & > & 350\kmps\,,\\
r_t & \gsim & 150\kpc\,. \\
\end{array}
\label{rho_vdisp_rt_values}
\end{equation}

Whereas we will have to await further astronomical observations to yield a 
complete, unbiased data set on the dSphs, it is expected that the above 
estimates, being conservative, will hold. The above estimates are also 
consistent with the earlier estimation of these
parameters by \citet{CRB96} using only the rotation curve data of the
Galaxy up to $R\lsim 20\kpc$. The current set of parameters yields a 
total mass of the Galaxy, including its DM halo, $M_{\rm
Galaxy}\sim 2\times10^{12}\msun$.
\section {Summary and Conclusions
\label{sec:summary}}
The theoretical density distribution of the dark matter particles 
constituting the halo depends sensitively on the competition between their 
velocity dispersion and the total gravitational potential generated by 
the dark and the baryonic matter of the Galaxy. We have exploited this 
sensitivity to estimate the parameters related to the phase space 
distribution of the dark matter in the Galaxy. This is carried out within 
the context of a self-consistent model of the dark matter halo which 
solves the collisionless Boltzmann and Poisson equations, in which the 
density distribution of baryonic matter is taken directly to fit the 
observations, while a specific functional form --- namely, the King or the 
truncated isothermal model --- for the phase space distribution 
of the dark matter particles in the halo is used. This DM halo 
model has three free parameters --- the density ($\rhodmsun$) 
and velocity dispersion ($\vdispdmsun$) in the solar neighborhood and the 
radius ($r_t$) of the halo. We have estimated these 
three parameters by comparing the theoretical 
predictions with the astronomical observations, namely, (a) the rotation 
curve of the Galaxy measured up to $\sim20\kpc$ and (b) the distance and 
radial velocities of the dwarf spheroidal satellites which probe the 
galactic gravitational field up to very large distances. Indeed, the 
dSphs, located as they are at distances well beyond $20\kpc$ with a 
broad peak around $\sim 100\kpc$, provide crucial inputs in fixing the 
parameters $\vdispdmsun$ and $r_t$ of the halo model. We have shown 
that the observed paucity of the dSphs at short Galactocentric distances  
imposes constraints on their possible orbits and 
makes their velocity distribution asymmetrical favoring transverse 
velocities over radial velocities. A special version of the virial theorem 
is then applied to the observed radial velocities of the dSphs to estimate 
the circular rotation speed at galactocentric distances of $\sim 100\kpc$. 
The self-consistent model reproduces the rotation speed and the observed 
distance distribution of the dSphs. Whereas one will 
have to await further astronomical observations which will yield rotation 
curves with less scatter and a more complete sample of dSphs, the 
available astronomical data indicates $\rhodmsun\approx 0.3\gev/\cm^3$, 
$\vdispdmsun > 350\kmps$ and $r_t\gsim 150\kpc$ for our assumed model of 
the halo. Improvements on these model parameters could also result from an 
analysis of the tidal streams surrounding the Milky Way system. 

Finally, we emphasize that while the estimates of (or constraints on) the 
DM halo parameters given above have been obtained within the context of a 
specific assumed model of the phase space distribution function of the DM 
particles in the Galaxy, namely the King model, we believe that 
other models for the phase space DF that are parametrizable in terms of 
the three coarse-grained parameters as above, and which give similar fits 
to the observational data on the rotation curve and the number 
distribution of the dSphs as obtained in this paper, 
will yield similar values of the three parameters. This, however, remains 
to be explicitly demonstrated. 
It will thus be interesting to repeat the analysis with
various other possible forms of the DF for the DM particles in order to
assess the robustness of the parameter values or the constraints on them
obtained in this paper for the DM halo of the Galaxy.

%
\section*{Acknowledgment}
\vspace*{-7mm}It is a pleasure to thank Ms.~G.~Rajalakshmi and
Mr.~Suresh Doravari for discussions. 
One of us (P.B.) would like to acknowledge the hospitality at the
Physics Department, Washington University and the Indo-US forum for 
travel support.

\newpage 
\appendix 
\begin{center}
 {\bf APPENDICES}
\end{center}
\section[A]{Rotation curves for various Galaxy model parameters 
\label{apndx:model_rotcurves}}
\begin{figure}
\resizebox{\textwidth}{!}{\rotatebox{270}{\includegraphics[]{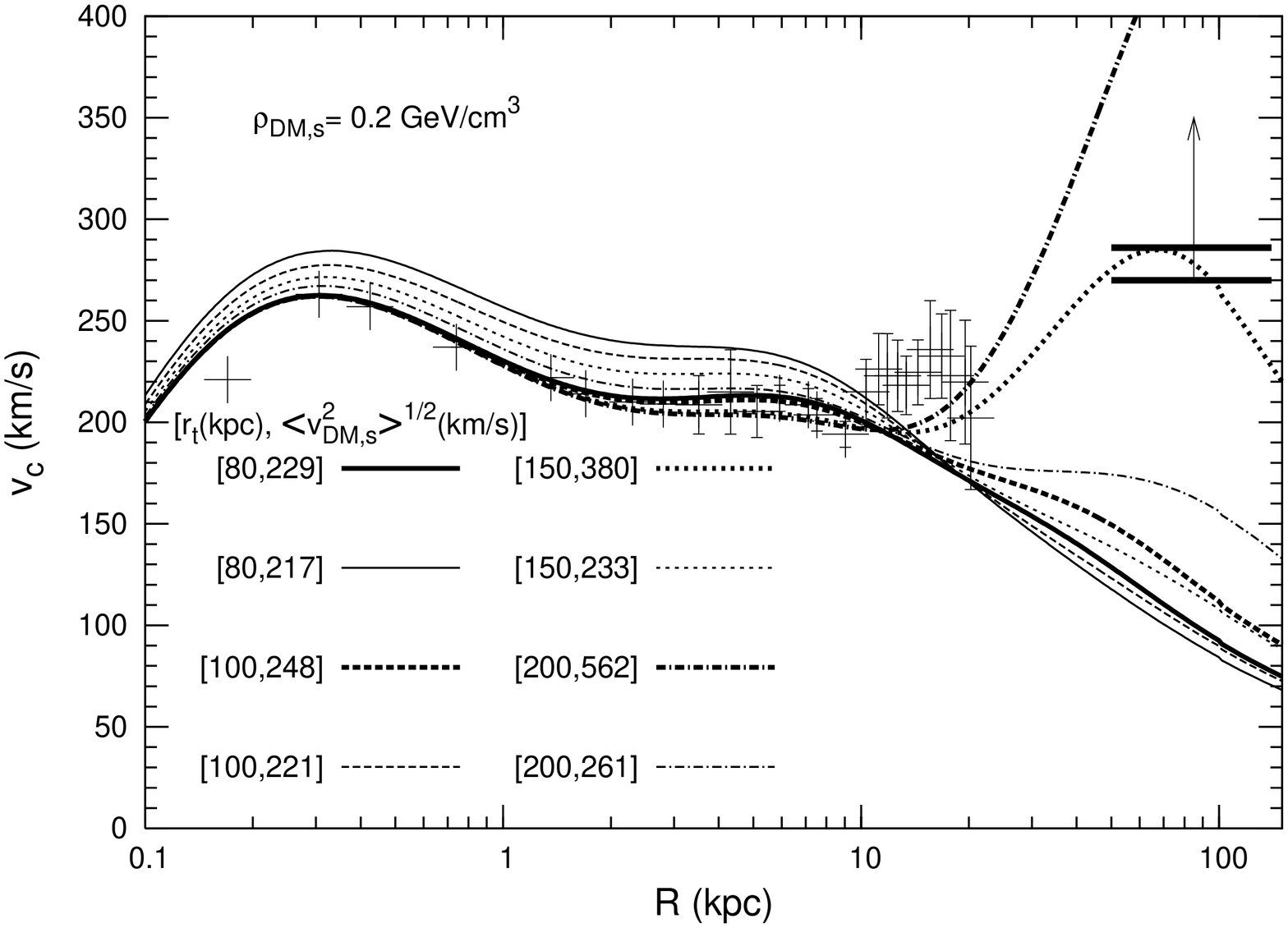}}} 
\caption{
Theoretically calculated rotation curves of the Galaxy based on the 
self-consistent model described in the main text, with 
$\rhodmsun=0.2\gev/\cm^3$ and values of $\vdispdmsun$ and $r_t$  
as indicated. (In the legends inside the Figure, the subscript $s$ is used 
in place of $\odot$). For each value of $r_t$ we show two RCs: the one 
with the thicker line corresponds to the maximum possible value of 
$\vdispdmsun$ consistent with the chosen values of $\rhodmsun$ and $r_t$, 
and the other with a lower value of $\vdispdmsun$, in order to indicate 
the 
possible range of the RCs obtained as one varies the values of 
$\vdispdmsun$ for 
fixed values of $\rhodmsun$ and $r_t$. The observational data are from 
\citet{HS97} for the case 
$R_0=8\kpc$ and $v_c(R=R_0)=200\kmps$. The two solid horizontal lines  
represent two different estimates of the expected lower limits on 
$v_c$ in the region of $R\sim 100\kpc$ obtained in this paper from the 
dynamics of the dwarf-spheroidals. 
The small glitches in the curves at $R\simeq100\kpc$ are artifacts of 
numerical calculation and are due to increased grid spacing used for 
distances $r\geq 100\kpc$) in order to reduce the total computation time.
} 
\label{fig:appendix_rc_fig_0.2}
\vspace{4in}
\end{figure}

\begin{figure}
\resizebox{\textwidth}{!}{\rotatebox{270}{\includegraphics[]{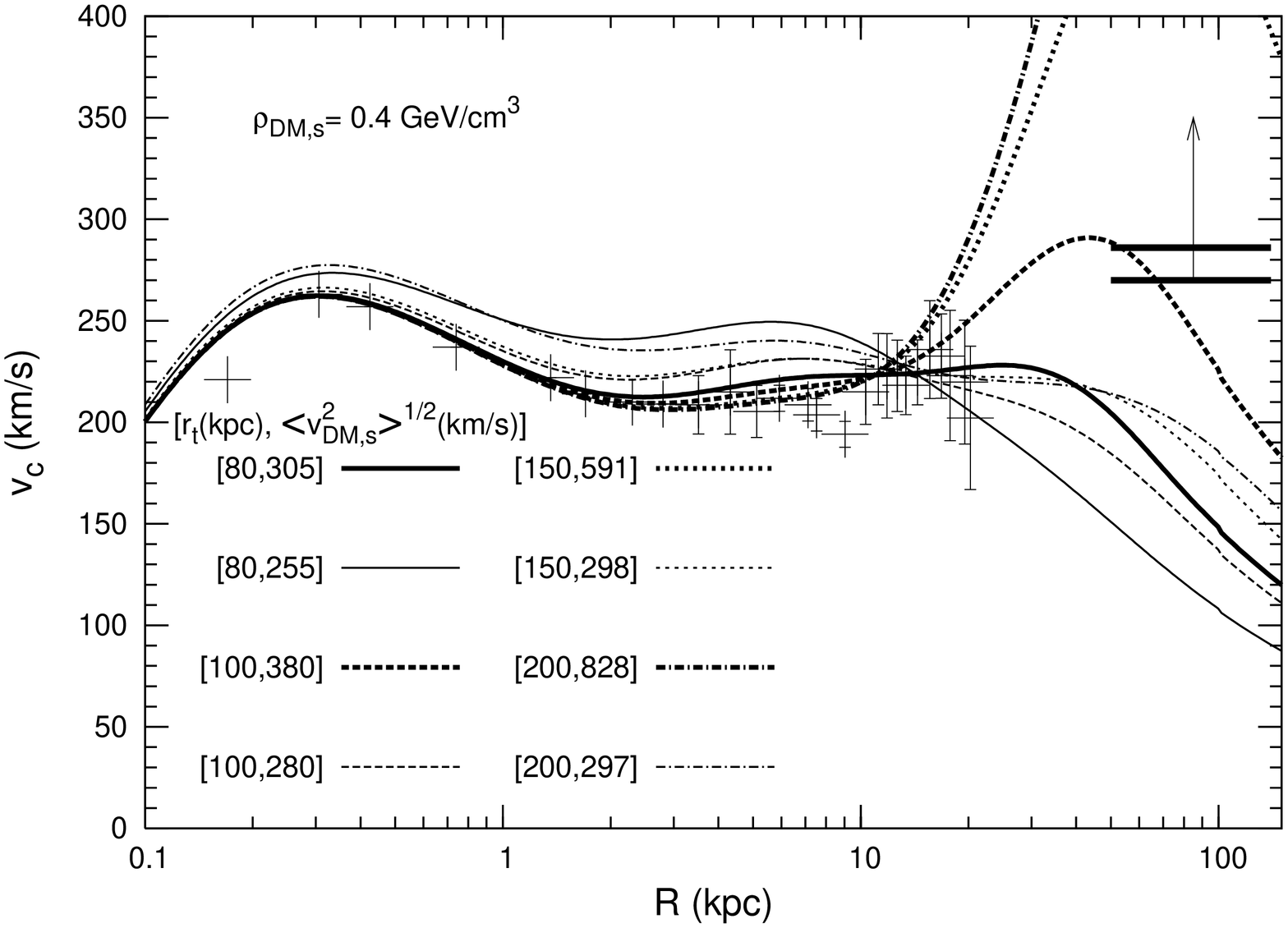}}} 
\caption{
Same as Fig.\ref{fig:appendix_rc_fig_0.2}, but for 
$\rhodmsun=0.4\gev/\cm^3$.
} 
\label{fig:appendix_rc_fig_0.4}
\vspace{5in}
\end{figure}

In this Appendix we display several sets of theoretical Rotation 
Curves (RCs) for the Galaxy covering a wider range (than what was given 
in the main body of the paper) of possible values of the Dark 
Matter (DM) halo parameters that describe our self-consistent model of the 
Galaxy used in this paper (see sections 
\ref{sec:dmhalo_model_math} and \ref{sec:rotation_curve}). These are shown 
in Figures \ref{fig:appendix_rc_fig_0.2} and 
\ref{fig:appendix_rc_fig_0.4}. 

The dark matter is 
modeled as having a lowered (truncated) isothermal distribution 
function [see equations (\ref{king_df}) and 
(\ref{relative_energy_def})] which is 
described by three parameters, namely, (i) the dark matter density in the 
solar neighborhood, $\rhodmsun$, (ii) the velocity dispersion in the 
solar neighborhood, $\vdispdmsun$, and (iii) the truncation radius, 
$r_t$.
%

\newpage 
\section[B]{Velocity asymmetry parameter for anisotropic 
distributions
\label{apndx:j_aniso_DF}}
In this Appendix we investigate the effect of deviations from an initially 
isotropic velocity distribution of the dSphs, on the velocity 
anisotropy parameter $j$ discussed in the main text. Specifically, we 
consider 
an initial distribution function (DF) that favors radial orbits and 
suppresses large transverse velocities. Such DFs can be obtained by 
introducing a dependence on the angular momentum $L$ which is an integral 
of motion in the relevant potential \citep{BT87}. Following the 
prescription in Binney and Tremaine let us consider, as an explicit 
example, an initially anisotropic DF for the dSphs obtained by multiplying 
the isothermal (and isotropic) DF (\ref{F_sat}) by a factor 
$L^{-2\alpha}$ with $0 < \alpha < 1$:
\begin{equation}
F_{\rm aniso}(\vt,\vr,r)\propto |L|^{-2\alpha} 
\exp\left\{-3\left(\frac{1}{2}(v_r^2+v_t^2)+\phi(r)\right)/\sigma^2_s\right\}\,,
\label{F_sat_aniso}
\end{equation} 
where $L=r\, \vt$. As before, $\sigma_s$ is the initial velocity 
dispersion of the satellites, and $\phi(r)$ is the gravitational potential 
in which the satellites move. The parameter $\alpha$ controls the amount 
of initial suppression of transverse velocities, with $\alpha=0$ 
representing the isotropic case discussed in the main body of the paper. 
The 
singularity of the factor $L^{-2\alpha}$ at $r\to 0$ can be treated as 
usual by suitably softening the factor near the origin. 

The exercise now is to impose a lower cutoff on the radial 
coordinate at $\rmin$ and numerically calculate the velocity anisotropy 
parameter $j$ (as a function of $\rmin$) given by equation (\ref{j_def2}), 
using the DF (\ref{F_sat_aniso}), for various values of the parameter 
$\alpha$. 

\begin{figure}
\rotatebox{270}{\includegraphics[width=4in]{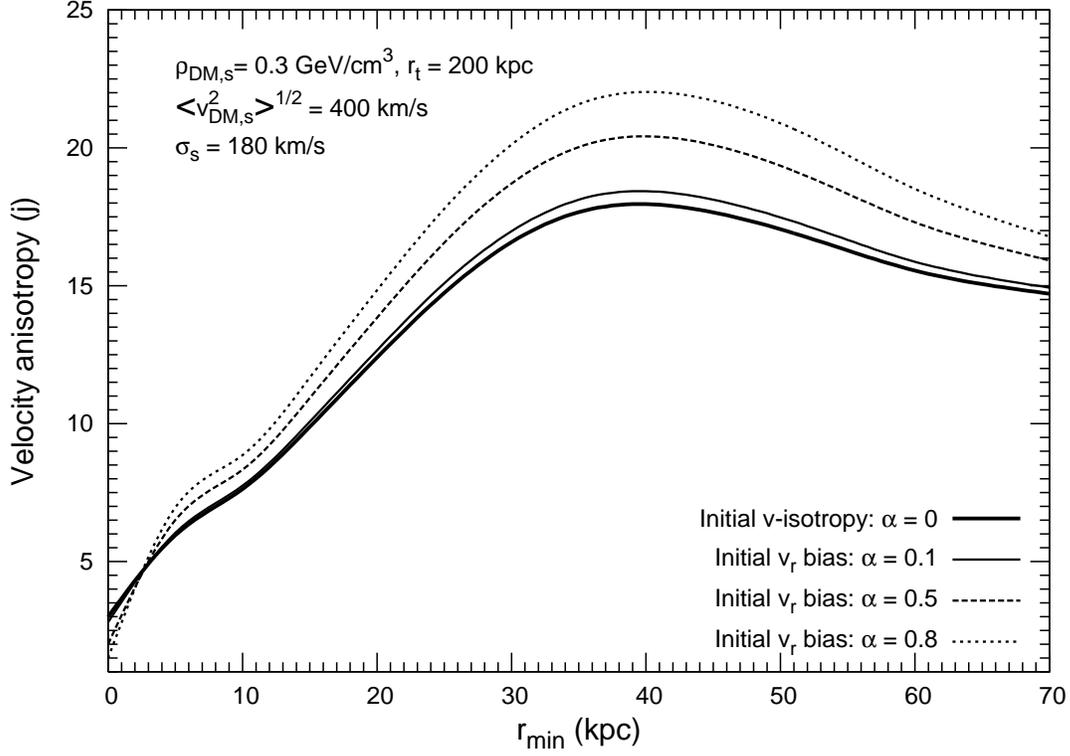}} 
\caption{
The velocity anisotropy parameter $j$ as a function of $\rmin$ 
for an initially anisotropic DF of the form $F_{\rm aniso} \propto 
|L|^{-2\alpha}\exp\left\{-3E/\sigma_s^2\right\}$ for $\sigma_s=180\kmps$ 
and 
four different values of $\alpha$ = 0, 0.1, 0.5, and 0.8, as indicated. 
The values of other relevant parameters are as indicated.   
} 
\label{fig:j_vrbias_p_alpha_ds180_dm200_400}
\end{figure}

\begin{figure}
\rotatebox{270}{\includegraphics[width=4in]{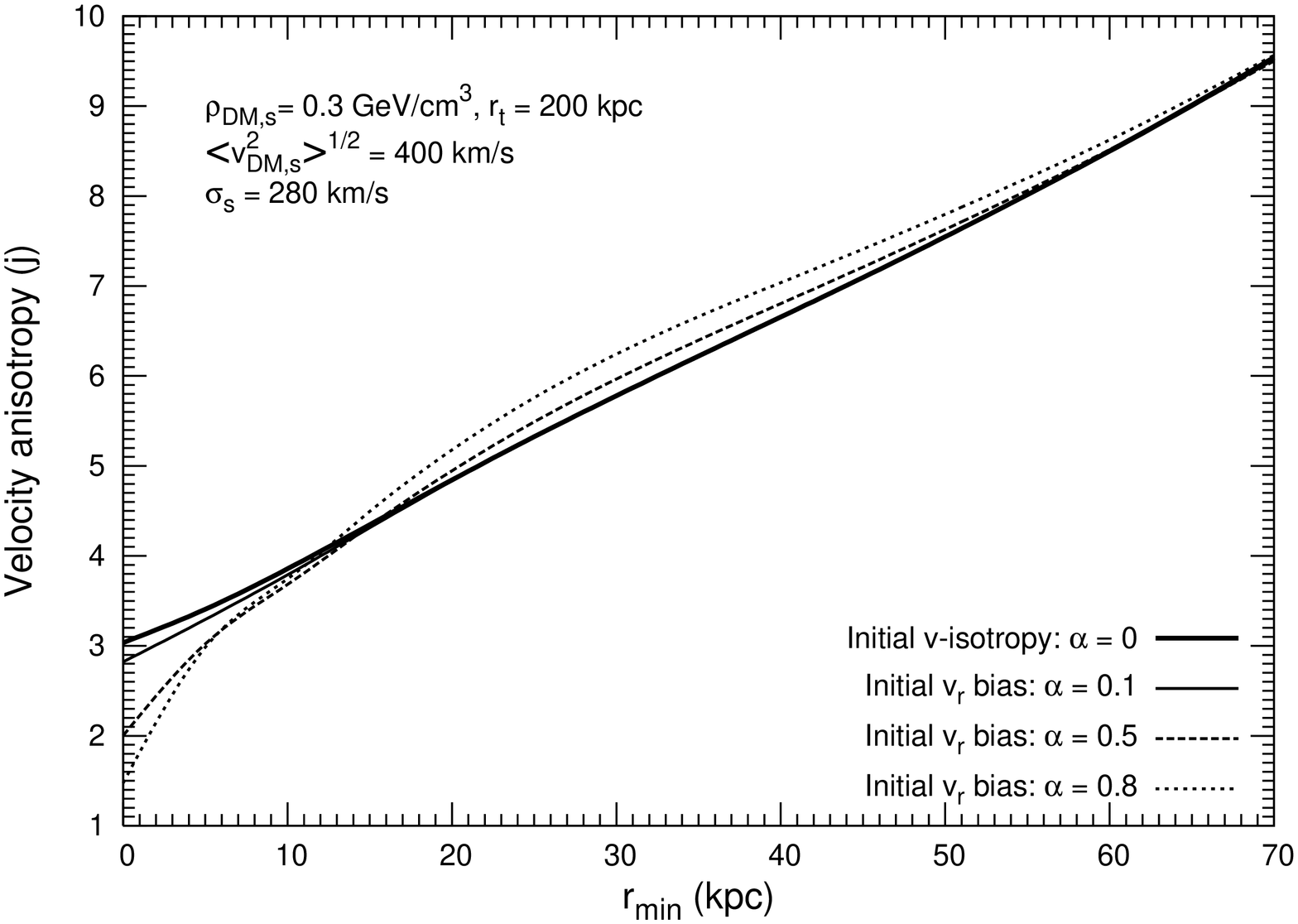}} 
\caption{
Same as Fig.~\ref{fig:j_vrbias_p_alpha_ds180_dm200_400} but for 
$\sigma_s=280\kmps$. 
} 
\label{fig:j_vrbias_p_alpha_ds280_dm200_400}
\vspace{5in}
\end{figure}

The results are shown in Figures 
\ref{fig:j_vrbias_p_alpha_ds180_dm200_400} 
and 
\ref{fig:j_vrbias_p_alpha_ds280_dm200_400} 
for $\sigma_s = 180\kmps\,$ and $280\kmps\,$,  
respectively, for example. 

From these Figures we see that the effect of the initial radial anisotropy 
survives only for small values of $\rmin$. As $\rmin$ increases, the value 
of $j$ increases rapidly with $\rmin$, the increase being {\it steeper} 
for larger values of the radial bias parameter $\alpha$.  

The above behavior of $j$ is easy to understand, when we note that as we 
try to suppress large values of $\vt$ by suppressing large values of 
$L (=r\vt)$, we are suppressing large values of $r$ as well. Stronger the 
suppression of large $L$, less is the probability of having orbits 
with large apogalacticons, $r_x$. Consequently, when orbits 
having perigalacticons, $r_n$, less than $\rmin$ are removed, we are 
left with a set of orbits with relatively higher values of the ratio 
$\eta=r_n/r_x$ compared to the case when there is 
no large $r$ suppression. These are nearly circular orbits, 
for which transverse velocities dominate over radial velocities (see 
Appendix C for a discussion of $j$ in terms of elliptical orbits), leading 
to increasingly larger values of $j$ as a function of $\rmin$ compared to 
the case when there is no large $L$ (and consequently large $r$) 
suppression. 

From the above discussions it is clear that, for the relevant ranges of 
values of the various parameters, the 
values of $j$ obtained for an initially radially biased DF are, in fact, 
{\it larger} than those for the initially isotropic DF. The values of $j$ 
obtained in the main body of the paper assuming an initially isotropic DF 
are thus conservative estimates.

\newpage 
\section[C]{Estimation of $j$-parameter from the kinematics of 
an elliptical orbit 
\label{apndx:j_ellipse}}
The orbits of dSphs in the Galaxy, in general, will follow rosette like 
patterns, which may not close. But for the present purposes let us 
approximate them as ellipses with various values of apo- and 
peri-galacticons, $r_x$ and $r_n$, but confined within the limits $\rmax$ 
and $\rmin$, respectively, as defined in the main text.  

We begin by writing the equation of an ellipse in polar coordinates 
$r, \psi$ (with origin at one of the foci): 

\begin{equation}
r=\frac{\ell}{1+\varepsilon\cos\psi}\,,
\label{ellipse_eqn}
\end{equation}
where $\ell$ is the semi-latus rectum and $\varepsilon$ the 
eccentricity; in terms of $r_x$ and $r_n$ these are given by 
\begin{equation}
\ell=\frac{2\eta r_x}{1+\eta}\,, \,\,\, 
\varepsilon=\frac{1-\eta}{1+\eta}\,, \,\,\, {\rm with} \,\,\,\, 
\eta=\frac{r_n}{r_x}\,.
\label{ellipse_alpha_ecc_eta_def}
\end{equation}
Note, also, that 
\begin{equation}
\ell=r_n(1+\varepsilon)=r_x(1-\varepsilon)\,. 
\end{equation}
Since the orbits of the dSphs sample large $r$ values in the halo, 
compared to the radial scale of the disk, the potential is nearly 
spherically symmetric and we may assume an approximate conservation of 
angular momentum $L$ \citep{BT87} and write, for the tranverse component 
of the velocity, 
\begin{equation}
v_\psi=r\frac{d\psi}{dt}=\frac{L}{r}\,.
\label{ellipse_v_psi}
\end{equation}
Now, the probability $P(r)dr$ of finding the dSph between $r$ and $r+dr$ 
is inversely proportional to the radial component of its velocity, $v_r$, 
at $r(\psi)$:
\begin{equation}
P(r)\propto \frac{1}{|v_r\left[r(\psi)\right]|}\,,
\label{ellipse_P(r)}
\end{equation} 
where 
\begin{equation}
v_r\left[r(\psi)\right] 
= \frac{dr}{dt}=\frac{dr}{d\psi}\frac{d\psi}{dt}=\frac{L\varepsilon 
\sin\psi}{\ell}\,.
\label{ellipse_v_r}
\end{equation}

With the probability $P(r)$ given by equation (\ref{ellipse_P(r)}) above, 
the velocity anisotropy parameter $j$ is given by 
\begin{equation}
j \equiv 1+\frac{\langle v_\psi^2 \rangle}{\langle v_r^2 \rangle}
 = 1 + \frac{\pi}{\varepsilon^2}\frac{1}{I_1}\,,
\label{ellipse_j_eqn}
\end{equation}
where $\langle v_\psi^2 \rangle$ and $\langle v_r^2\rangle$ are the mean 
square transverse and radial velocities, respectively, and 
\begin{equation}
I_1=\int_0^\pi 
\frac{\sin^2\psi}{\left(1+\varepsilon\cos\psi\right)^2} d\psi\,.
\label{ellipse_I1_integral}
\end{equation}

We can also calculate the average location of the satellite on the 
ellipse, $r_{\rm av}\equiv \int r P(r) dr /\int P(r) dr$, in terms of 
either $r_x$ or $r_n$. This gives 
\begin{equation}
r_{\rm av}=\ell \frac{I_2}{I_3}\,,
\label{ellipse_rav}
\end{equation}
where 
\begin{equation}
I_2=\int_0^\pi 
\frac{d\psi}{\left(1+\varepsilon\cos\psi\right)^3}\,, \,\,\, {\rm 
and}\,\,\,\, 
I_3=\int_0^\pi 
\frac{d\psi}{\left(1+\varepsilon\cos\psi\right)^2}\,. 
\label{ellipse_I2andI3_integrals}
\end{equation}

\begin{figure}
\rotatebox{270}{\includegraphics[width=4in]{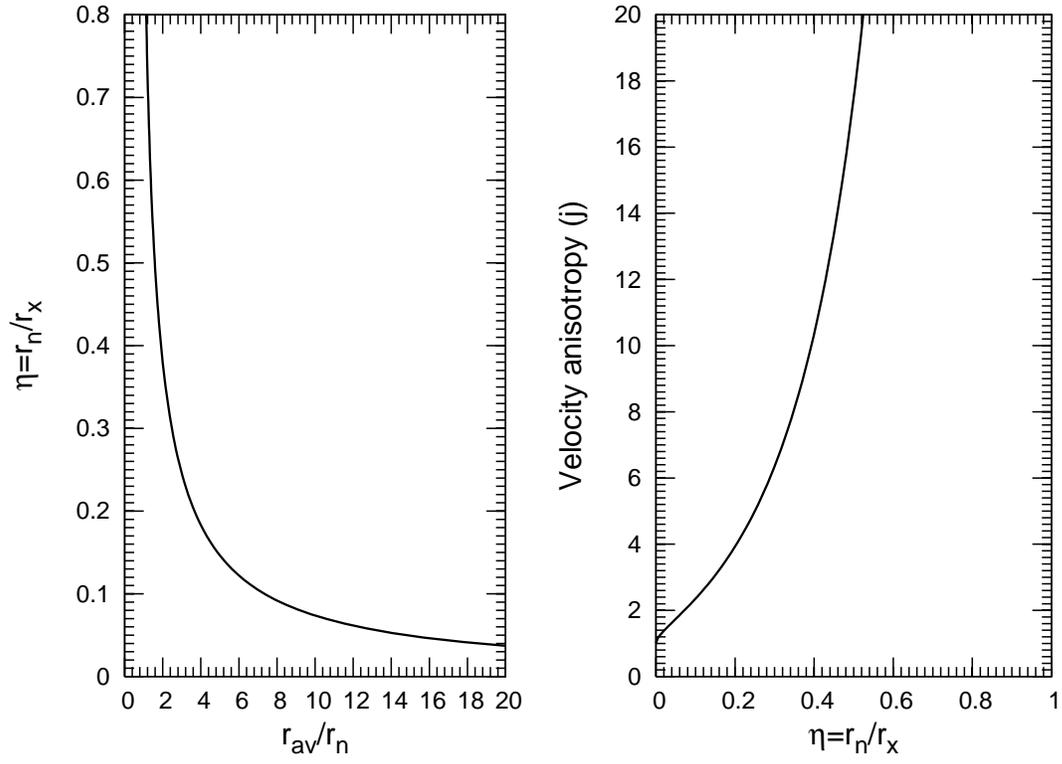}}
\caption{The ratio $\eta=r_n/r_x$ {\it vs.} the average location, $r_{\rm 
av}$ (left panel), and the velocity anisotropy parameter $j$ {\it vs.} 
$\eta$ (right panel), for a particle in an elliptical orbit.}   
\label{fig:ellipse_j_rav}
\vspace{4in}
\end{figure}

The integrals $I_1$, $I_2$ and $I_3$ are easily evaluated numerically. The 
resulting values of $j$ and $r_{\rm av}$ for the ellipse under 
consideration, as functions of the parameter $\eta=r_n/r_x$, are displayed 
in Figure \ref{fig:ellipse_j_rav}. 

The observed value of $r_{\rm av}$ for the mean location of the dSphs 
estimated from Figure \ref{fig:hist_dsph_CML} is $\approx 150\kpc$ 
which, with our most likely value of $\rmin\approx 60\kpc$, gives $r_{\rm 
av}/\rmin \lsim 2.5$, which from the left panel in Figure 
\ref{fig:ellipse_j_rav} corresponds to $\eta=r_n/r_x\gsim 0.3$. For 
this value of $\eta$, 
then, we can read off the value of $j$ from the left panel of Figure 
\ref{fig:ellipse_j_rav}, giving $j\gsim 6.4$. 

The analysis presented in this Appendix thus gives a 
kinematic insight into the values of $j$ parameter derived from a 
consideration of the phase space distribution in the main body of the 
paper. 


\end{document}